\documentclass[]{article}

\usepackage{graphicx}
\usepackage[round]{natbib}
\usepackage{multirow}
\usepackage{booktabs}
\usepackage{amsmath}
\usepackage{amssymb}
\usepackage{amsfonts}
\usepackage[dvipsnames]{xcolor}
\usepackage{moreverb}
\usepackage{multirow}
\usepackage[geometry]{ifsym}
\usepackage{MnSymbol}
\usepackage{pifont}
\usepackage{lpic}
\usepackage{fullpage}

\def\tm{\leavevmode\hbox{$\rm {}^{TM}$}}

\title{The energy cascade in grid-generated non-equilibrium decaying turbulence}
\author{P. C. Valente and J. C. Vassilicos \\ 
Department of Aeronautics, Imperial College London,
London SW7 2AZ, United Kingdom }

\begin{document}

\maketitle

\begin{abstract}
We investigate non-equilibrium turbulence where the non-dimensionalised dissipation coefficient
$C_{\varepsilon}$ scales as $C_{\varepsilon} \sim Re_{M}^{m}/Re_{\ell}^{n}$ with
 $m\approx 1 \approx n$ ($Re_M$ and $Re_{\ell}$ are global/inlet and local Reynolds numbers respectively) by measuring the downstream evolution of the scale-by-scale energy transfer, dissipation, advection, production and transport in the lee of a square-mesh grid and compare with a region of equilibrium turbulence (i.e. where  $C_{\varepsilon}\approx \mathrm{constant}$).
These are the main terms of the inhomogeneous, anisotropic version of the von K\'{a}rm\'{a}n-Howarth-Monin equation.
It is shown in the grid-generated turbulence studied here that, even in the presence of non-negligible turbulence production and transport, production and transport are
large-scale phenomena that do not contribute to the scale-by-scale balance for scales smaller than about a third of the integral-length scale, $\ell$, and therefore do not affect the energy transfer to the small-scales.
In both the non-equilibrium and the equilibrium decay regions, the peak of the scale-by-scale energy transfer scales as $(\overline{u^2})^{3/2}/\ell$ ($\overline{u^2}$ is the variance of the longitudinal fluctuating velocity).
In the non-equilibrium case this scaling implies an imbalance between the energy transfer to the small scales and the dissipation.
This imbalance is reflected on the small-scale advection which becomes larger in proportion to the maximum energy transfer as the turbulence decays whereas it stays proportionally constant in the further downstream equilibrium region where $C_{\varepsilon} \approx \mathrm{constant}$ even though $Re_{\ell}$ is lower.
\end{abstract}

\section{Introduction}

Recent work on fractal and regular grid-generated decaying turbulence showed that there are two distinct turbulence decay regions demarcated by two different behaviours of the kinetic energy dissipation per unit
volume, $\varepsilon$, at high Reynolds numbers (see \cite[see][and references therein]{VV2012}. 
A non-equilibrium region closer to the grid where $C_{\varepsilon} \equiv \varepsilon \ell/u'^3 \sim Re_M^m/Re_{\ell}^n$  (with $m\approx n\approx 1$ for the highest Reynolds number data), the local Reynolds number is high and the energy spectrum has a power law shape over a wide range of wavenubers with exponent close to $-5/3$; and an equilibrium region further downstream where the Reynolds number has
dropped but $C_{\varepsilon} \sim \mathrm{constant}$ ($u'$ and $\ell$ are, respectively, the root-mean-square of the fluctuating velocity and an integral length-scale; $Re_{\ell} = u' \ell/\nu$ and $Re_M = U_{\infty} M/\nu$ with $U_{\infty}$ being the inlet velocity, $M$ an inlet mesh size and $\nu$ the kinematic viscosity). 
This dichotomy of dissipation behaviours, and in particular the new non-equilibrium dissipation scalings, have
been found with different measurement techniques and by different research groups \cite[see e.g.][]{MV2010,gomesfernandesetal12,discettietal11,Nagata2012}. 

In this paper we attempt to flesh out the meaning and some of the properties of non-equilibrium turbulence and what distinguishes it from equilibrium turbulence where $C_{\varepsilon} \sim \mathrm{constant}$. 
In particular we want to investigate the  connection between the non-equilibrium dissipation behaviour and the behaviour of the inertial energy cascade flux, $\Pi$ ($C_{\Pi} \equiv \Pi \ell/u'^{3}$ in dimensionless terms).
We do this on the basis of two-point two-component measurements which allow the estimation of nearly all terms in the inhomogeneous and anisotropic form of the von K\'{a}rm\'{a}n-Howarth-Monin equation \cite[see e.g.][]{Danaila2012}. 
These terms represent turbulent dissipation and scale-by-scale transport, production, advection, energy transfer and viscous diffusion.

We chose to carry out this study in the lee of two regular grids, RG60 and RG115 (see \citealt{VV2012}), for three reasons: (i) the new non-equilibrium dissipation law is most clearly defined in the lee of our regular grids; (ii) the usual equilibrium $C_{\varepsilon} \sim \mathrm{constant}$ law is accessible in our wind tunnels only with RG60; and (iii) the same test section region can be used to study non-equilibrium turbulence with RG115 and equilibrium turbulence with RG60 (see \S \ref{sec:3}).

\subsection{Balance between energy cascade and dissipation (Kolmogorov's $4/5^{\mathrm{th}}$ law)} \label{sec:45law}

In a seminal contribution, \cite{K41c} arrived to an exact expression (i.e. without adjustable constants) relating the  third-order structure function ($\overline{\delta u_{\parallel}^3}$) and the dissipation ($\varepsilon$) within the inertial-range.
The starting point for the derivation is the von K\'{a}rm\'{a}n-Howarth equation \cite{KH1938} simplified using the framework put forward in his earlier work \cite[]{K41a}. 
The expression, $\overline{\delta u_{\parallel}^3}=-4/5\,\varepsilon\, r$, is commonly known as  Kolmogorov's $4/5^{\mathrm{th}}$ law due to the pre-factor appearing in the equation which follows from the hypothesis of local isotropy ($\delta u_{\parallel}(r)$ is the component of the velocity difference parallel to the separation vector $\mathbf{r}$ and $r=|\mathbf{r}|$). 
Note that one can relax the local isotropy constraint by averaging the third-order structure function over all solid angles and arrive to a more general ``$4/5^{\mathrm{th}}$-type law'' \cite[see][]{NT99}. 
Since the third-order structure function is readily interpreted as a scale-by-scale (non-linear)
energy flux, we refer to general  ``$4/5^{\mathrm{th}}$-type laws'' as  $\Pi-\varepsilon$ balance, where $\Pi$ denotes the energy transfer in the inertial range ($\Pi = -5/4\,d\,\overline{\delta u_{\parallel}^3}/dr$ in Kolmogorov's $4/5^{\mathrm{th}}$-law).
This balance and the related $4/5^{\mathrm{th}}$-type laws represent the essence of the Richardson-Kolmogorov cascade.

Its importance can hardly be overstated since it serves as a basis of many theories and models of turbulence. This is readily identified whenever the dynamical role of the inertial range is characterised by a single scalar quantity, i.e. $\varepsilon$ (including Kolmogorov's earlier work). Richardson's pair diffusion law as well as theories of acceleration-, pressure-, passive and active scalar-field characteristics in the inertial range \cite[]{MY75} are all examples of the implicit use of the dissipation as a measure of the instantaneous energy transfer across the inertial range and permitting a phenomenological theory to be constructed. 
Other related examples of the use of the $\Pi-\varepsilon$ balance, with some modifications, can be found in theories of polymer drag reduction \cite[see e.g.][]{deGennes:book} and magnetohydrodynamics \cite[see e.g.][]{Biskamp:book}.

For homogeneous stationary turbulence the $\Pi-\varepsilon$ balance can be derived rigorously \cite[assuming finiteness of the dissipation for vanishing viscosity or a related theoretical limit, see \citealt{Frisch:book} and][]{NT99} and has received substantial experimental and numerical support \cite[see e.g.][]{Antonia2006}. 
However, the merit of Kolmogorov's work is the formulation of a theory for ``the case of an arbitrary turbulent flow with sufficiently large Reynolds number'' \cite[]{K41a} by introducing hypotheses of local homogeneity, local isotropy and local stationarity. 
(Note that by local stationarity we are referring to Kolmogorov's idea that ``within short time intervals [the small scale fluctuations] can naturally be regarded as being stationary, even when the flow as a whole is non-stationary'', \citealp{K41a} -- see also \citealp{George2013} for a critique of this concept.) 
So far, in the case of temporally or spatially evolving turbulent flows the support of the $\Pi-\varepsilon$ balance is still meagre, arguably because the current laboratory and numerical experiments do not reach a sufficiently large Reynolds number for the onset of an inertial range \cite[$Re_{\lambda}\gtrsim \mathcal{O}(10^6)$ according to][]{Antonia2006}. 
Still, in the above mentioned literature there is a latent expectation that a $\Pi-\varepsilon$ balance will hold at extremely high Reynolds numbers and the departures are broadly denoted as `finite Reynolds number' (FRN) effects  \cite[]{Qian99,Moisy1999,Lundgren2002,Lundgren2003,Gagne2004,Antonia2006,Cambon2012}. 

In contrast with the above viewpoint, one can find literature (typically pertaining to turbulence modelling) questioning the validity of the instantaneous balance between energy transfer and dissipation in non-stationary and in spatially evolving flows  \cite[]{Schiestel87,Lumley92,Yoshizawa1994,RB2009} and advocating the necessity to account for the transfer time of kinetic energy from large to small scales \cite[]{Lumley92,Schiestel87}. 
In fact, there is no local nor instantaneous balance between energy transfer and dissipation even in statistically stationary and homogeneous turbulence as pointed out by  \cite{Kraichnan74} and subsequently evidenced in direct numerical simulations by \cite{BO98}. However, this balance does nevertheless hold on average in statistically stationary and homogeneous turbulence if the Reynolds number is high enough but it does not in time-evolving (e.g. decaying) or spatially-developing turbulence where the transfer time of kinetic
energy from large to small scales, i.e. the time-lag, therefore becomes critically important in the description of the turbulence cascade \cite[]{Schiestel87,Lumley92, Yoshizawa1994,BO98,Bos2007,RB2009,MY2010}.

The time-lag and non-equilibrium theories of \cite{Schiestel87,Lumley92,Yoshizawa1994, Bos2007} (among others) can, in principle, be applied throughout the decay region of grid-generated turbulence and therefore over both the first part of the decay region where $C_{\varepsilon} \sim Re_{M}^{m}/Re_{\ell}^{n}$ with $m\approx 1 \approx n$ and the second, further downstream and lower $Re_{\ell}$ part, where $C_{\varepsilon} \approx \mathrm{constant}$. These theories, at least as they currently stand, can therefore not explain the new dissipation law $C_{\varepsilon} \sim Re_{M}^{m}/Re_{\ell}^{n}$ and have not predicted
it. 
In fact, some forms of these theories \cite[see][]{Bos2007} predict $C_{\varepsilon} \approx \mathrm{constant}$ for decaying turbulence but with a higher constant value of $C_{\varepsilon}$ than for forced statistically stationary turbulence. 
The increased constant value of $C_{\varepsilon}$ is a consequence of the cascade time-lag.

We therefore distinguish between non-equilibrium decaying turbulence where $C_{\varepsilon} \sim Re_{M}^{m}/Re_{\ell}^{n}$ and the time-lag non-equilibrium turbulence of \cite{Schiestel87,Lumley92,Yoshizawa1994, Bos2007}. 
This does not mean that there is no cascade time-lag in non-equilibrium decaying turbulence where $C_{\varepsilon} \sim Re_{M}^{m}/Re_{\ell}^{n}$, it simply means that this time-lag is not sufficient by itself to explain this new type of non-equilibrium. 
The far downstream relatively lower Reynolds number grid-generated turbulence which we refer to as equilibrium turbulence and where $C_{\varepsilon} \approx \mathrm{constant}$ may in fact be no more than a time-lag non-equilibrium turbulence as in \cite{Schiestel87,Lumley92,Yoshizawa1994, Bos2007}. 
It is important to keep in mind the different meanings of the terms equilibrium and non-equilibrium according to context to avoid confusion. 
We now proceed with the inhomogeneous and anisotropic form of the von K\'{a}rm\'{a}n-Howarth-Monin equation \cite[see][]{Danaila2012} which forms the basis of the present
study.

\subsection{Scale-by-scale energy transfer budget equation} \label{sec:KHM}

A scale-by-scale energy transfer budget similar to the von K\'{a}rm\'{a}n-Howarth-Monin equation \cite[see (22.15) in][]{MY75}, but extended to inhomogeneous turbulent flows, can be derived directly from the Navier-Stokes \cite[see e.g.][ and references therein]{Deissler61,Casciola04,Danaila2012}.

The starting point is the incompressible Navier-Stokes decomposed into mean and fluctuating components at two  distinct locations $\mathbf{x} \equiv \mathbf{X} + \mathbf{r}/2$ and $\mathbf{x'} \equiv \mathbf{X} - \mathbf{r}/2$ ($\mathbf{X}$ is the centroid of the two points and $r= |\mathbf{r}|$ their distance),
\begin{equation}
\left\lbrace
\begin{aligned}
\frac{\partial \,U_i + u_i}{\partial t} + U_k \frac{\partial u_i}{\partial x_k} + u_k \frac{\partial U_i}{\partial x_k} + U_k \frac{\partial U_i}{\partial x_k}  + u_k \frac{\partial u_i}{\partial x_k} = & -\frac{1}{\rho}\frac{\partial \,P + p}{\partial x_i} \,\,+ \nu \,\frac{\partial^2 \,U_i + u_i}{\partial x_k^2 }\\
\frac{\partial \,U'_i + u'_i}{\partial t} + U'_k \frac{\partial u'_i}{\partial x'_k} + u'_k \frac{\partial U'_i}{\partial x'_k} + U'_k \frac{\partial U'_i}{\partial x'_k} + u'_k \frac{\partial u'_i}{\partial x'_k} = & -\frac{1}{\rho}\frac{\partial \,P' + p'}{\partial x'_i} + \nu \frac{\partial^2\, U'_i + u'_i}{\partial x'^2_k},
\end{aligned}
\right.
\end{equation}
together with the continuity equations ($\partial U_k/ \partial x_k = \partial U'_k/ \partial x'_k =\partial u_k/ \partial x_k = \partial u'_k/ \partial x'_k = 0$). In the present notation  $U_i\equiv U_i(\mathbf{x})$, $u_i\equiv u_i(\mathbf{x})$, $P\equiv P(\mathbf{x})$, $U'_i\equiv U_i(\mathbf{x'})$, $u'_i\equiv u_i(\mathbf{x'})$ and $P'\equiv P(\mathbf{x'})$. 

The main steps in the derivation are to (i) subtract the two equations above and denote the velocity differences as $\delta u_i \equiv u_i - u'_i$, $\delta p \equiv p - p'$ and $\delta U_i \equiv U_i - U'_i$, (ii) multiply the resulting expression by $2\delta u_i$, (iii) ensemble average over an infinite number of realisations (denoted by overbars; in practice ergodicity is used on the basis of the time
stationarity at a given point in our spatially evolving flows and time averages are performed) and (iv) change the coordinate system from ($\mathbf{x}$, $\mathbf{x'}$) to ($\mathbf{X}$, $\mathbf{r}$).
The resulting equation reads,
\begin{equation}
\begin{aligned}
\frac{\partial \, \overline{\delta q^2}}{\partial t} +  
\left( \frac{U_k + U'_k}{2} \right) \frac{\partial \, \overline{\delta q^2}}{\partial X_k} &+ 
\frac{\partial \, \overline{\delta u_k \delta q^2}}{\partial r_k} +
\frac{\partial \, \delta U_k \overline{\delta q^2}}{\partial r_k}  = \\
-2\overline{\delta u_i \delta u_k} \frac{\partial \, \delta U_i}{\partial r_k} &-
\overline{(u_k + u'_k)\delta u_i}\, \frac{\partial \, \delta U_i}{\partial X_k} -
\frac{\partial}{\partial X_k} \left( \overline{\frac{(u_k + u'_k) \delta q^2}{2}}  \right) -\\
\frac{2}{\rho} \frac{\partial \, \overline{\delta u_k \delta p} }{\partial X_k}  &+
\nu \left[ 2\frac{\partial^2 }{\partial r^2_k} +  \frac{1}{2}\frac{\partial^2 }{\partial X^2_k}  \right] \overline{\delta q^2}  - 
 2\nu \left[ \overline{ \left(  \frac{\partial u_i}{\partial x_k}  \right)^2} + \overline{ \left(  \frac{\partial u'_i}{\partial x'_k}  \right)^2}\right],
\end{aligned}
\label{eq:KHM}
\end{equation}
 where $\overline{\delta q^2} \equiv \overline{(\delta u_i)^2}$ (with summation over the index $i=1,2,3$). 
 Equation \eqref{eq:KHM} is essentially an inhomogeneous von K\'{a}rm\'{a}n-Howarth-Monin equation with additional terms to account for the inhomogeneity of the turbulent flow field. 
Each of the terms can be interpreted as follows. 
\begin{enumerate}
\item  $4\mathcal{A}_{t}^*(\mathbf{X},\mathbf{r})\equiv \partial \, \overline{\delta q^2}/\partial t$ results from the time dependence that $\overline{\delta q^2}(\mathbf{X},\,\mathbf{r})$ can have in certain unsteady flows.

\item  $4\mathcal{A}^*(\mathbf{X},\mathbf{r}) \equiv (U_k + U'_k)/2\,\, \partial \, \overline{\delta q^2}/\partial X_k$ represents an advection contribution to the change of  $\overline{\delta q^2}(\mathbf{X},\,\mathbf{r})$.

\item $4\Pi^*(\mathbf{X},\mathbf{r}) \equiv\partial \, \overline{\delta u_k \delta q^2}/\partial r_k$ represents a contribution which relates to nonlinear transfer of energy from the orientation point $\mathbf{r}/r$ on a spherical shell of radius $r$ centred at $\mathbf{X}$ to (a) concentric shells of larger radii (effectively to smaller radii since this term is typically negative) and (b) to other orientations within the same spherical shell. 
Notice that $\Pi^*$ is the divergence with respect to  $\mathbf{r}$ of the flux  $\overline{\delta u_k\delta q^2}$ and that owing to Gauss's theorem, $\iiint_{|\mathbf{r}|\leq r}\! \Pi^*\,dV = \oiint_{|\mathbf{r}|=r} \overline{\delta \mathbf{u} \delta q^2}\cdot\mathbf{r}/r \,dS$, i.e. the net contribution of $\Pi^*$ integrated over the sphere $|\mathbf{r}|\leq r$ is equal to the total radial flux over the spherical shell $|\mathbf{r}|=r$.
If the turbulence is homogeneous the radial flux is zero in the limit $r\rightarrow \infty$ and $4\Pi^*$ is indeed, unequivocally, a transfer term. 
Also note that (using a spherical coordinate system $(r,\,\theta,\,\phi)$ for $\mathbf{r}$) the integrals of the polar, $ \Pi^*_{\theta} $, and azimuthal, $\Pi^*_{\phi}$, contributions to the divergence $\Pi^*$ over the solid angle $\mathbf{r}/r$ are identically zero,  $\oiint_{|\mathbf{r}|=r} \Pi^*_{\theta} \,dS=\oiint_{|\mathbf{r}|=r} \Pi^*_{\phi} \,dS =0$, thus indicating a role of $\Pi^*$ in redistributing energy within a spherical shell.  

\item $4\Pi^*_U(\mathbf{X},\mathbf{r}) \equiv \partial \, \delta U_k\overline{\delta q^2}/\partial r_k$ represents  a contribution which relates to linear transfer of energy by mean velocity gradients from the orientation point $\mathbf{r}/r$ on a spherical shell of radius $r$ centred at $\mathbf{X}$ to concentric shells of larger radii. The motivation for this interpretation is analogous to that given for $\Pi^*$, where the turbulent flux is now $\delta U_k\overline{\delta q^2}$ \cite[see also][where the physical  interpretation of this term is given in wavenumber space]{Deissler61,Deissler81}.

\item $4\mathcal{P}^*(\mathbf{X},\mathbf{r}) \equiv-2\overline{\delta u_i \delta u_k} \, \partial \, \delta U_i / \partial r_k-\overline{(u_k + u'_k)\delta u_i}\, \partial \, \delta U_i/\partial X_k$ represents a contribution which relates to turbulent production. 
It is easiest to identify $\mathcal{P}^*$  as a production term by writing it in ($\mathbf{x}$, $\mathbf{x'}$)  coordinates, i.e.  $2\mathcal{P}^*=-\overline{u_i  u_k} \, \partial \,  U_i / \partial x_k - \overline{u'_i  u'_k} \, \partial \,  U'_i / \partial x'_k + \overline{u_i  u'_k} \, \partial \,  U_i / \partial x_k + \overline{u_i  u'_k} \, \partial \,  U'_i / \partial x'_k$, and recognising that the first two terms on the right-hand side  are the usual production terms of the single-point turbulent kinetic energy transport equation evaluated at $\mathbf{x}$ and $\mathbf{x'}$, respectively.

\item $4\mathcal{T}^*(\mathbf{X},\mathbf{r}) \equiv -\partial/\partial X_k \left(\overline{(u_k + u'_k) \delta q^2}/2+2/\rho\,\overline{ \delta u_k\delta p}\right)$ represents scale-by-scale turbulent transport from the orientation point $\mathbf{r}/r$ on a spherical shell of radius $r$ centred at $\mathbf{X}$  to an adjacent  shell (centred at $\mathbf{X} + \delta \mathbf{X}$) with the same radius and at the same orientation. Notice that $\mathcal{T}^*$ is the divergence with respect to  $\mathbf{X}$ of the flux  $-\overline{(u_k + u'_k) \delta q^2}/2-2/\rho\,\overline{ \delta u_k\delta p}$ and thus, making use of Gauss's theorem, it follows that the net contribution of $\mathcal{T}^*$ integrated (with respect to $\mathbf{X}$ for each $\mathbf{r}$) over a volume $V$ is equal to the total flux over the bounding surface of $V$. This motivates the physical interpretation of this term as a scale-by-scale turbulent transport. 

\item $4\mathcal{D}^*_{\nu} (\mathbf{X},\mathbf{r}) \equiv 2\nu\,\partial^2 \overline{\delta q^2}/\partial r^2_k$ represents  viscous diffusion around the orientation point $\mathbf{r}/r$ on a spherical shell of radius $r$ centred at $\mathbf{X}$ (note that $\lim_{r\rightarrow 0} \mathcal{D}^*_\nu (\mathbf{X},\mathbf{r}) = \varepsilon (\mathbf{X})$).

\item $4\mathcal{D}^*_{X,\nu} (\mathbf{X},\mathbf{r}) \equiv\nu/2\,\partial^2 \overline{\delta q^2}/\partial X^2_k$ represents scale-by-scale transport via viscous diffusion around the orientation point $\mathbf{r}/r$ on a spherical shell of radius $r$ centred at $\mathbf{X}$. This can be seen as a transport term following the same reasoning  as that made for $\mathcal{T}^*$ by noticing that  $4\mathcal{D}^*_{X,\nu}$ can be written as a divergence of the viscous flux $\nu/2\,\partial \overline{\delta q^2}/\partial X_k$.

\item $4\varepsilon^*(\mathbf{X},\mathbf{r}) \equiv 2\nu \overline{ \left( \partial u_i/\partial x_k  \right)^2} + 2\nu\overline{ \left( \partial u'_i/\partial x'_k  \right)^2}$ represents the sum of twice the turbulent kinetic energy dissipation at the two locations, i.e. $2\varepsilon + 2\varepsilon' = 4\varepsilon^*$ with $\varepsilon^*\equiv (\varepsilon + \varepsilon')/2$, 
where $\varepsilon = \nu \overline{ \left( \partial u_i/\partial x_k  \right)^2}$ and $\varepsilon' = \nu\overline{ \left( \partial u'_i/\partial x'_k  \right)^2}$. \\

\end{enumerate}

For large enough $r$,  \eqref{eq:KHM} reduces to four times the average of two single-point turbulent kinetic energy transport equations, one evaluated at $\mathbf{x}$ and the other at $\mathbf{x'}$ \cite[see][]{Casciola04}.
Recall that the dependence on the orientation $\mathbf{r}/r$ can be removed by averaging the terms over spherical shells of radius $r$, in the spirit of  \cite{NT99}.
The spherical shell averaged terms are denoted by removing the superscript asterisk.

\subsection{Outline} \label{sec:outline}
This paper is organised as follows. In \S 2 the details of the experimental apparatus are presented together with all the necessary  \emph{a priori} checks to ensure the quality of the collected data. In \S 3 we specify how each of the terms in \eqref{eq:KHM} is estimated from the data and discuss the downstream variation in the anisotropy of the two-point second- and third-order structure functions. In \S 4 we discuss the role of turbulence production and transport on the other terms in \eqref{eq:KHM}. In \S 5 we discuss the scaling of the scale-by-scale energy transfer, advection and viscous diffusion as the flow decays for both the non-equilibrium and the equilibrium dissipation regions and summarize the main findings in \S 6.


\section{Experimental setup}
\subsection{Measurement apparatus} \label{sec:apparatus}
The experiments are performed in a 0.46 m x 0.46 m x 3.5 m blow-down wind tunnel at the Department of Aeronautics  in Imperial College London \cite[for further details see][]{VV2011,VV2012}.

\begin{figure}
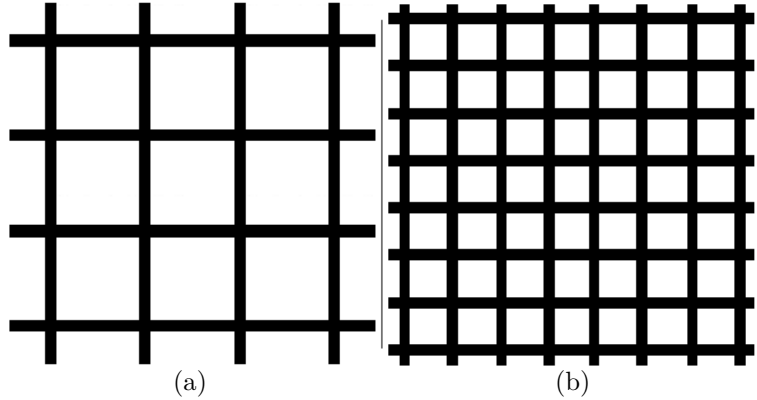

   \centering 
   \begin{lpic}{Figures/GridsRG115RG60(100mm)}
   \lbl{45,-4;(a)}
   \lbl{140,-4;(b)}
   \end{lpic}
   \vspace{3mm}
\caption{Sketch of the turbulence generating grids, (a) RG115 and (b) RG60. The RG115 is a mono-planar regular grid with a distance between parallel bars of $M=115\mathrm{mm}$, lateral and longitudinal bar thicknesses of $t_0=10\mathrm{mm}$ and $d=3.2\mathrm{mm}$, respectively and a blockage ratio of $\sigma = 17\%$. The RG60 is a bi-planar regular grid with $M=60\mathrm{mm}$, $t_0=d=10\mathrm{mm}$ and $\sigma = 32\%$. Owing to the geometrical differences between the RG115 and RG60 the location of the turbulent kinetic energy peak along the centreline ($y=z=0$) is considerably different: $x_{\mathrm{peak}}\approx 0.83\mathrm{m}$ and $x_{\mathrm{peak}}\approx 0.14 \mathrm{m}$, respectively. }
\label{fig:grids}
\end{figure} 

The measurement apparatus to compute estimates of the terms in \eqref{eq:KHM} (except the pressure transport term) consists of two X-probes (aligned with the xy plane to measure the longitudinal and vertical velocity components, $U$ and $V$ for the mean and $u$, $v$ for the fluctuating components) mounted on a traverse mechanism controlling the vertical distance between the probes and their individual pitch angle for \it in-situ \rm calibration.
(Note that, in the orthonormal coordinate system used, $x$ is aligned with the mean flow and $y$ \& $z$ are perpendicular and parallel to the floor, respectively.)
Data are acquired in the lee of two regular grids,  RG115 and RG60, sketched in figure \ref{fig:grids} and described in the caption of the figure. 

This apparatus was previously used  to measure two transverse velocity correlation functions in \cite{VV2013} (henceforth referred to as \textbf{I}) where a detailed description of the traverse and measurement system can be found together with an assessment of the measurement resolution and mutual interference between the two X-probes.
For convenience we recall that one of the vertical traverse systems displaces the two probes symmetrically about their centroid (defined as the geometrical midpoint between the X-probes' centres) whereas the second vertical traverse system displaces the centroid keeping the separation between the probes fixed. For short probe separations the distance between the X-probes is optically measured with an external camera which is used to set the reference separation as well as to ensure high position accuracy. The minimum vertical separation between the probes is $\Delta y =1.2$mm (probe resolution $\sim 4\eta - 8\eta$) whereas the maximum separation in the measurements is $70$mm ($\sim 2 L$). In total $23$ separations are measured ($\Delta y=1.2,$ $1.6,$ $2.0$, $2.5,$ $3.0,$ $3.5,$ $4,$ $5,$ $6,$  $8,$  $10,$ $12,$ $14,$ $16,$ $20,$ $24,$ $28,$ $32,$ $36,$ $44,$ $52,$ $60,$ $70$mm).  
It was found that the overall precision of the prescribed vertical separation between the X-probes was typically $\pm 50\mu$m (i.e. over the three degrees of freedom: vertical symmetric displacement and pitching of the two individual probes used for calibration).

\subsection{Data acquisition and statistical convergence} \label{sec:conv}
The in-built signal conditioners of the anemometer are set to analogically filter at  $30$kHz and to offset and amplify the signal $-1$V  and $2\times$, respectively. 
The four analogue anemometer signals are sampled at 62.5kHz with a National Instruments NI-6229 (USB) with a resolution of 16-bit over a range of $[-1\,\,1]$V.  
The turbulent velocity signals are acquired for $9$min corresponding to $150\,000-200\,000$ integral-time scales.
The data acquisition, wind tunnel speed and traverse motors control are performed with MATLAB\tm. 

Perhaps the most demanding statistic of interest here, in terms of statistical convergence, is the triple structure function $\overline{\delta u_i \delta q^2}$.
To quantify its statistical uncertainty, we assign 95\% confidence intervals to the measurements ($\pm 1.96 \sqrt{\mathrm{var}(\overline{\delta u_i \delta q^2})}$; see \citealt{BG96}). 
The sampling variance ($\mathrm{var}(\overline{\delta u_i \delta q^2})$) is estimated as
\begin{equation}
\mathrm{var}(\delta u_i \delta q^2) = \frac{1}{N}\left(\overline{(\delta u_i \delta q^2)^2} - \overline{\delta u_i \delta q^2}^2\right),
\label{eq:vardudq2}
\end{equation}
where $N$ is the number of independent samples. 
(Note that $\overline{\delta u_i \delta q^2}$ is non-central statistical moment, hence equation 4 of \cite{BG96}, which is derived for central moments, has additional uncertainty terms which are not applicable, see \cite{KS58} for further details.) 

\begin{figure}
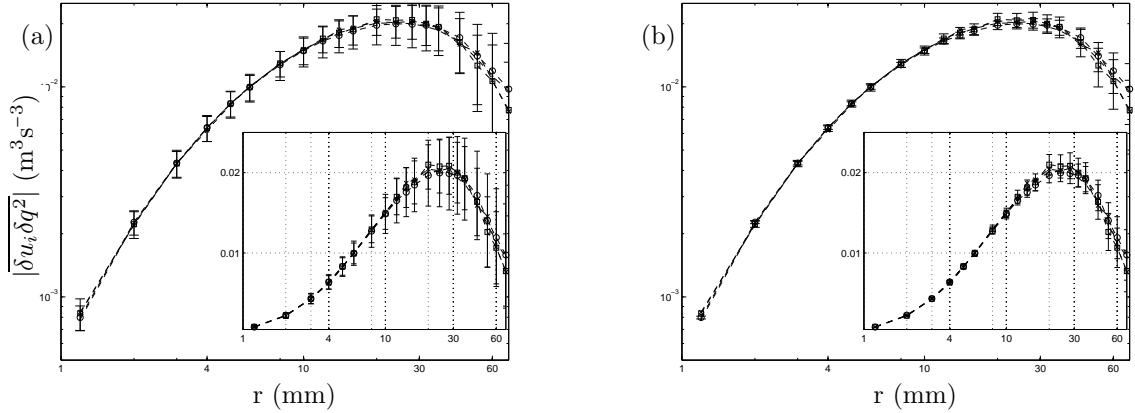

\centering
\begin{minipage}[c]{0.5\linewidth}
   \centering
   \begin{lpic}{Figures/ErrorBarsI(67mm)}
   \lbl{9,125;(a)}
   \lbl[W]{4,75,90;$|\overline{\delta u_i \delta q^2}|$ (m$^3$s$^{-3}$)}
   \lbl[W]{100,0;r (mm)}
   \end{lpic}
\end{minipage}%
\begin{minipage}[c]{0.5\linewidth}
   \centering 
   \begin{lpic}{Figures/ErrorBarsII(67mm)}
   \lbl{9,125;(b)}
   \lbl[W]{5,75,90;\hspace{8mm}${}^{}$\hspace{10mm}}
   \lbl[W]{100,0;r (mm)}
   \end{lpic}
\end{minipage}
\caption[Confidence intervals of $|\overline{\delta u_i \delta q^2}|$]{$|\overline{\delta u_i \delta q^2}|$ versus (\protect\raisebox{-0.5ex}{\SmallSquare} $\!|\!\!$ \protect\raisebox{-0.5ex}{\SmallCross}) longitudinal separations $(r_x,\,r_y)=(r,\,0)$ and (\protect\raisebox{-0.5ex}{\SmallCircle} $\!|\!\!$ \protect\raisebox{0ex}{\tiny{\FilledSmallCircle}}) transverse separations $(r_x,\,r_y)=(0,\,r)$; $95\%$ confidence intervals estimated with (a) global integral time-scale and (b) tailored integral time-scale characteristic of $\overline{\delta u_i \delta q^2}(r_x,\,r_y)$. Note that the same experiment is repeated twice and (\protect\raisebox{-0.5ex}{\SmallSquare} $\!|\!\!$ \protect\raisebox{-0.5ex}{\SmallCircle}) represent exp. I and  (\protect\raisebox{-0.5ex}{\SmallCross} $\!|\!\!$ \protect\raisebox{0ex}{\tiny{\FilledSmallCircle}}) exp. II.}
\label{fig:ErrorBars}
\end{figure}

The repeatability of the measurement of $\overline{\delta u_i \delta q^2}$ was also assessed in a precursory experiment by repeating the same measurement twice (data acquired at the lee of RG60 for the centroid at $(x,y,z)=(1250,\,0,\,0)$mm).
Comparing the repeatability (a somewhat more stringent test) with the estimated uncertainty (figure \ref{fig:ErrorBars}a), we notice that the confidence intervals are excessively large.
This could indicate that the number of independent samples, $N$, based on the integral-time scale, is underestimated. 
Indeed, splitting the data into integral-time scale sized blocks and extracting a single sample of $\delta u_i \delta q^2(r_x,\,r_y)$, leads to estimates of $\overline{\delta u_i \delta q^2}(r_x,\,r_y)$ with significantly more scatter, indicating that uncorrelated samples were lost. 
Instead of using the standard integral time-scale, we can define alternative de-correlation time-scales by taking the autocorrelation of $\delta u_i \delta q^2(r_x,\,r_y)$ at two times with varying lags and then integrating the resulting correlation functions. 
This methodology provides a tailored integral time-scale representative of the de-correlation length associated with  $\overline{\delta u_i \delta q^2}$ at each $(r_x,\,r_y)$. 
Assuming that twice this tailored integral time-scale is the characteristic lag between independent samples of  $\delta u_i \delta q^2(r_x,\,r_y)$ we get new estimates of $N(r_x,\,r_y)$, and consequently new confidence intervals, which are shown in figure \ref{fig:ErrorBars}b. 
Note that, experiments performed with Particle Imaging Velocimetry obtain reasonable estimates of $\overline{\delta u_i \delta q^2}$ \cite[see e.g.][]{Moisy2011,Danaila2012}  even though the number of independent samples is $\mathcal{O}(10^3)$.

The error bars of the spherically averaged divergence of $\overline{\delta u_i \delta q^2}$ (figures \ref{fig:inhomo}-\ref{fig:PeakValues}) include the $95\%$ confidence intervals plus the error due to the uncertainty of the vertical separation between the X-probe $\approx \pm 50\mu$m (see \S \ref{sec:apparatus}). The two uncertainties are stacked with a standard propagation of error formula applied to the central differences scheme.

\section{Description of the experimental results} \label{sec:3}

The data are acquired with $\mathbf{X}$, the midpoint between the two X-probes,  along the centreline ($y=z=0$) at five downstream locations between  $x=1250$mm and $x=3050$mm
($X_1=1250,\,1700,\,2150,\,2600,\,3050\mathrm{mm}$ and $X_2=X_3=0$).
For two downstream locations of the centroid, $X_1=1250$mm and $X_1=2150$mm, additional datasets off-centreline at $X_2=-6$mm and $X_3=0$ are acquired so that derivatives of the statistics with respect to $X_2$ can be computed, particularly those needed to estimate $\partial/\partial X_2 \,\overline{(v + v') \delta q^2}$, see \eqref{eq:KHM}. 
The choice of  6mm as the distance to evaluate the $X_2$-derivative is based on the single-point data in the lee of RG115-turbulence used in \textbf{I} to estimate the lateral triple-correlation transport (i.e. $\partial/\partial y \,\overline{v  q^2}$). Based on those data it is found that  the spanwise derivative $\partial/\partial y \,\overline{v  q^2}$ is well approximated by $(\overline{v  q^2}(h_y)-\overline{v  q^2}(0))/h_y$ up to spacings of $h_y\approx 8$mm. Too small $h_y$ introduce unnecessary uncertainty to the estimates.

Recall that the X-probes are symmetrically traversed in the y-direction with respect to a fixed $\mathbf{X}$, thus enabling the measurement of the statistical correlations as a function of $r_2$, and that the dependence on $r_1$ is recovered using Taylor's hypothesis. 
On the other hand, the traverse mechanism does not allow displacements in the z-direction and the measurements are restricted to the  vertical xy-plane at $z=0$ and thus $X_3=0$ and $r_3=0$.

The downstream range of the measurements corresponds to $8-21x_{\mathrm{peak}}$ for RG60 and $1.5-3.7x_{\mathrm{peak}}$ for RG115, a stark difference in the streamwise range relative to $x_{\mathrm{peak}}$ owing to the geometrical differences between the grids (see figure \ref{fig:grids}). 
In effect, the measurement range for the RG115 corresponds to a nonclassical energy dissipation region whereas for the RG60 it corresponds to a classical one (see \citealt{VV2012}), thus allowing their direct comparison. 
Recall that the $Re_{\lambda}$ ranges,  as well as the straddled Kolmogorov microscales $\eta$, are comparable for both grids at the chosen measurement locations. Hence the same set up can be used in both experiments without penalising resolution ($84 \leq Re_{\lambda}\leq 100$ versus $105 \leq Re_{\lambda}\leq 140$ and $0.19\mathrm{mm} \leq \eta \leq 0.32\mathrm{mm}$ versus $0.16\mathrm{mm} \leq \eta \leq 0.28\mathrm{mm}$ for RG60- and  RG115-generated turbulence, respectively).

Note that the Taylor microscale is calculated as $\lambda = (15 \nu\, \overline{u^2}/\varepsilon)^{1/2}$ and the Kolmogorov microscale as $\eta = (\nu^3/\varepsilon)^{1/4}$.
The dissipation rate is estimated as $\varepsilon =\varepsilon^{\mathrm{iso,3}} \equiv \nu(\overline{(\partial u/\partial x)^2}+2\overline{(\partial v/\partial x)^2}+4\overline{(\partial u/\partial y)^2}+2\overline{(\partial v/\partial y)^2})$ (see discussion in \S 5 of \textbf{I}).

\begin{table}
\centering
\begin{tabular}{rcccccc}
Location & $1250$ & $1700$ & $2150$ & $2600$ & $3050$ \\
$x/x_{\mathrm{peak}}$ & $8.5$ & $11.5$ & $15.6$ & $17.6$ & $20.7$ \\
$Re_{\lambda}$ & $100$ & $94$ & $89$ & $87$ & $84$ \\
$\overline{u^2}\,(\mathrm{m^2s^{-2}})$  & $0.15$ & $0.10$ & $0.07$ & $0.06$ & $0.05$ \\
$\lambda \, (\mathrm{mm})$ & $3.8$ & $4.4$ & $5.0$ & $5.5$ & $5.8$ \\
$\eta\, (\mathrm{mm})$ & $0.19$ & $0.23$ & $0.27$ & $0.30$ & $0.32$ \\
$\varepsilon\,(\mathrm{m^{2}s^{-3}})$& $2.51\pm0.07$ &  $1.18\pm0.02$ & $0.66\pm0.02$ & $0.42\pm0.01$ & $0.30\pm0.01$ \\
\end{tabular} 
\caption{Turbulence statistics for the RG60}
\label{tab:RG60}
\end{table} 
\begin{table}
\centering
\begin{tabular}{rcccccc}
\centering
Location & $1250$ & $1700$ & $2150$ & $2600$ & $3050$ \\
$x/x_{\mathrm{peak}}$ & $1.5$ & $2.0$ & $2.6$ & $3.1$ & $3.7$ \\
$Re_{\lambda}$ & $140$ & $126$ & $118$ & $110$ & $105$ \\
$\overline{u^2}\,(\mathrm{m^2s^{-2}})$  & $0.32$ & $0.20$ & $0.14$ & $0.10$ & $0.08$ \\
$\lambda \, (\mathrm{mm})$ & $3.7$ & $4.2$ & $4.7$ & $5.1$ & $5.6$ \\
$\eta\, (\mathrm{mm})$ & $0.16$ & $0.19$ & $0.22$ & $0.25$ & $0.28$ \\
$\varepsilon\,(\mathrm{m^{2}s^{-3}})$&  $5.21\pm0.24$ & $2.59\pm0.12$ & $1.48\pm0.02$ & $0.89\pm0.02$ & $0.55\pm0.04$\\
\end{tabular} 
\caption{Turbulence statistics for the RG115}
\label{tab:RG115}
\end{table}

\subsection{Estimation of the terms in the inhomogeneous  K\'{a}rm\'{a}n-Howarth-Monin equation}\label{sec:KHMcomputed}

We now describe how the terms appearing in the inhomogeneous  K\'{a}rm\'{a}n-Howarth-Monin equation \eqref{eq:KHM}  are estimated from the present two-component, two-dimensional data using the statistical  characteristics of the flow and some additional assumptions. 

From the spatially-varying two-component turbulent signals, acquired simultaneously at the $23$ transverse separations, the second- and third-order structure functions ($\overline{(\delta u)^2}$, $\overline{(\delta v)^2}$, $\overline{(\delta u)^3}$, $\overline{(\delta v)^3}$, $\overline{\delta u(\delta v)^2}$, $\overline{\delta v(\delta u)^2}$) and the mixed structure functions ($\overline{(v + v')\delta u}$, $\overline{(u + u') (\delta u)^2}$, $\overline{(v + v') (\delta u)^2}$, $\overline{(u + u') (\delta v)^2}$, $\overline{(v + v') (\delta v)^2}$) are computed for all $(r_1,\,r_2)$. 
(Note that $r_2$ are just the $23$ transverse separations ($1.2\mathrm{mm}\leq \Delta y \leq 70$mm) and $r_1=n_i\,f_s/U_{\infty}$  where $f_s/U_{\infty}$ is the spatial sampling frequency by virtue of Taylor's hypothesis and $n_i$ are 23 integers chosen to yield approximately the same separation magnitudes as those in the transverse direction.)

\begin{figure}
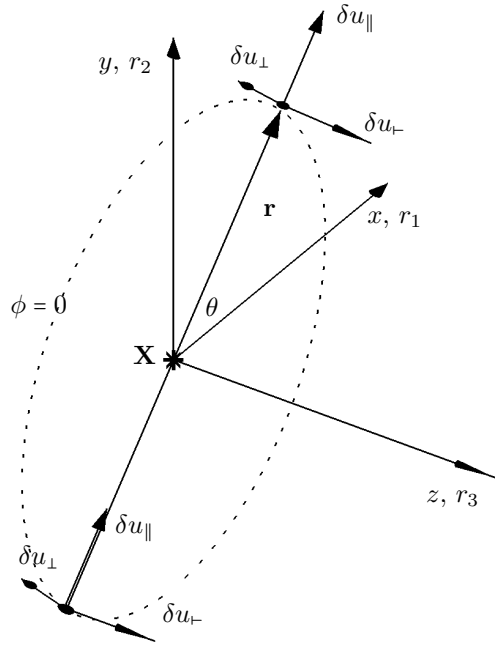

\centering
\begin{lpic}{Figures/Thesis_SketchAssumptions(65mm)}
   \lbl{39,44;$x$, $r_1$}
   \lbl{11,60;$y$, $r_2$}
   \lbl{45,15;$z$, $r_3$}
   \lbl{26,45;$\mathbf{r}$}
   \lbl{13,30;$\mathbf{X}$}
   \lbl{20,35;$\theta$}
   \lbl{2,35;$\phi=0$}
   \lbl{35,65;$\delta u_{\parallel}$}
   \lbl{24,61;$\delta u_{\perp}$}
   \lbl{38,54;$\delta u_{\vdash}$}
   \lbl{12,12;$\delta u_{\parallel}$}
   \lbl{2,9;$\delta u_{\perp}$}
   \lbl{17,3;  $\delta u_{\vdash}$}
\end{lpic}
\caption[Sketch of the three velocity-difference components, $\delta u_{\parallel}$, $\delta u_{\perp}$ and $\delta u_{\vdash}$]{Sketch of the three velocity-difference components ($\delta u_{\parallel}$, $\delta u_{\perp}$, $\delta u_{\vdash}$). $\delta u_{\parallel}$ and $\delta u_{\perp}$ are the velocity-difference components lying in the measurement plane ($r_3=0$) which are, respectively, parallel and perpendicular to the separation vector $\mathbf{r}$. $\delta u_{\vdash}$ is the velocity-difference component orthogonal to the other two ($\delta u_{\vdash}$). In the spherical coordinate system used here, $\theta$ is the angle between $\mathbf{r}$ and the $r_1$--axis and $\phi=0$ corresponds to the $r_1$--$r_2$ plane.}
\label{fig:sketch}
\end{figure}

The structure functions are then bi-linearly interpolated onto a spherical coordinate system $(r,\,\theta,\,\phi=0)$ such that $(r,\,0,\,0)$ is aligned with $r_1$ and $(r,\,\pi/2,\,0)$ with $r_2$ (see figure \ref{fig:sketch}). 
The grid points in the new coordinate system are located at the interceptions between the 23 circumferences of radius $r$ and $19$ equally spaced radial lines between the polar angles $\theta = [0\,\,\,\,\pi/2]$.
After the interpolation, the data is smoothed with a weighted average between each data point at $(r,\,\theta)$ and its neighbours $(r \pm \Delta r,\,\theta \pm \Delta \theta)$ (the total weight of the neighbouring points amounts to $37.5\%$). 

The greatest limitation of the present measurements is lacking the data for the third velocity component, $w$.
\cite{Moisy2011} faced the same limitation in their PIV data which they negotiated by considering the two-component surrogates  of the structure functions, which may be sufficient to make qualitative inferences.
However, the aim here is to obtain quantitative estimates of the terms in \eqref{eq:KHM}.
This is achieved by assuming inter-component axisymmetry of the velocity-difference statistics relative to the $\mathbf{r}$-axis. 
In other words, it is assumed  that the statistics of the two velocity-difference components perpendicular to $\mathbf{r}=(r_1,\,r_2,\,0)$ ($\delta u_{\perp}$ and $\delta u_{\vdash}$, see figure \ref{fig:sketch}) are approximately equal.
For the second-order structure function this assumption leads to $\overline{(\delta q)^2}(\mathbf{r})=\overline{(\delta u_{\parallel})^2}+2\overline{(\delta u_{\perp})^2}$. 
Similarly, for the third-order structure functions,  $\overline{\delta u_i(\delta q)^2}(\mathbf{r}) = \overline{\delta u_i(\delta u_{\parallel})^2}+2\overline{\delta u_i(\delta u_{\perp})^2}$ and  $\overline{(u_i + u'_i)(\delta q)^2}(\mathbf{r}) = \overline{(u_i + u'_i)(\delta u_{\parallel})^2}+2\overline{(u_i + u'_i)(\delta u_{\perp})^2}$.  
Note that this assumption is weaker than complete isotropy as it allows for dependence on the orientation $\mathbf{r}/r$.
Nevertheless, there is no presently available data to substantiate this assumption  and therefore its validity requires further investigation. 
Even so, it has been verified against all the present data that the added component (i.e. the factor 2 in the equalities two sentences above) does not change the qualitative behaviour of the structure functions, only their magnitude. 

Using the processed data, each of the terms in \eqref{eq:KHM}, except the pressure transport, is estimated at the measurement plane  as follows. 
Note that the numerical derivatives, both first and second order are computed using a three-point, non-equally spaced central differences scheme \cite[]{SB09}. For equally spaced derivatives this algorithm returns the usual standard central differences scheme. 

\begin{itemize}
\item  $\mathcal{A}_{t}^* = 0$ since grid-generated turbulence is stationary in the Eulerian frame.

\item  $4\mathcal{A}^* \approx\! (U + U')/2\,\partial /\partial  X_1\, \overline{\delta q^2}$ since the mean flow is approximately parallel, $V \approx W \approx 0$ and consequently,  the advection in the y- and z-directions is negligible.
The streamwise derivatives $\partial /\partial  X_1\,\overline{\delta q^2}$ are actually computed as  $\overline{\delta q^2}/X_1\,\,\partial /\partial \log X_1\, \left(\log\overline{\delta q^2}\right)$ using the three-point central differences scheme referenced above. 
Consequently, we use three datasets at different $X_1$ to compute the advection at each  $X_1$ for all $r$ (recall that there are a total of five datasets with $X_1=1250$, $1700$, $2150$, $2600$ and $3050$mm, respectively). 
Even though the various $X_1$ are coarsely spaced, it has been verified against the present data that the decay of $\overline{\delta q^2}$ for every $r$ can be reasonably approximated with a power-law whose virtual origin coincides with the location of the grid (just as a check of the formula used to compute streamwise
derivatives). 
Also, the longitudinal gradients of the mean velocity are small and therefore we make use of $(U+U')/2 \approx \left(U(X_1,\,X_2 + r_2 /2,\,0)+U(X_1,\,X_2 - r_2 /2,\,0)\right)/2$ to calculate $(U+U')/2 = \left(U(X_1+r_1/2,\,X_2 + r_2 /2,\,0)+U(X_1-r_1/2,\,X_2 - r_2 /2,\,0^{})\right)/2$. 

\item $4\Pi^*  \approx 1/r^2\,\partial/\partial r \left(r^2 \overline{\delta u_{\parallel} \delta q^2} \right)+ 1/(r\sin\theta)\,\partial /\partial \theta\left( \overline{\delta u_{\perp} \delta q^2}\right)$, i.e. the divergence is computed in the spherical coordinate system and the azimuthal component is assumed to be negligible owing to the axisymmetry of the turbulence statistics with respect to the centreline (see  discussion at the end of this subsection). Future work will be required to assess this assumption.

\item $\Pi^*_U\approx r_1\partial U /\partial x  \,\, \partial /\partial r_1\left(\overline{\delta q^2}\right)\approx 0$ see Appendix \ref{sec:neglect}.

\item $4\mathcal{P}^*  \approx 2\overline{(\delta u)^2} \,\partial U/\partial x + 4\overline{(v + v')\delta u}\, \partial U/\partial y$ since $V \approx W \approx 0$ and $\partial U/\partial z = \partial U'/\partial z'\approx0$ due to the expected symmetry of the mean flow relative to the plane $z=0$ which is parallel to the tunnel's vertical walls, includes the centreline and cuts the tunnel longitudinally in half. Also note that the symmetry of the mean flow relative to the centreline  (leading to $\partial U/\partial x \approx \partial U'/\partial x'$, $\partial U/\partial y \approx - \partial U'/\partial y'$) has been used to simplify $\partial\, \delta U/\partial r_k=1/2\left(\partial U/\partial x_k + \partial U'/\partial x'_k\right)$ as $\partial U/\partial x$ and $\partial\, \delta U/\partial X_k=\partial U/\partial x_k - \partial U'/\partial x'_k$ as $2\partial U/\partial y$.
 The transverse gradient $\partial U/\partial y$ is taken from a $12^{\mathrm{th}}$-order polynomial fit to the mean velocity data at each $X_1$ and the longitudinal gradient $\partial U/\partial x$ is computed as described in the previous item.

\item $4\mathcal{T}^* \approx -\partial/\partial X_1 \left(\overline{(u + u') \delta q^2}/2 \right)-\partial/\partial X_2 \left(\overline{(v + v') \delta q^2} \right) - 4\mathcal{T}_p^* $. 
The transverse derivative $\partial/\partial X_2(\overline{(v + v') \delta q^2}/2)$ ($\approx \partial/\partial X_3 \overline{(w + w')\delta q^2}/2$ owing to the symmetry of the turbulence statistics to $90^{\circ}$ rotations because of the grid's  geometry) is only computed where the additional off-centreline measurements  are acquired. 
The transverse derivative is simply taken as the difference between centreline and off-centreline data divided by their distance. 
The derivative with respect to $X_1$ is computed using the various datasets with different $X_1$. However, this can only be considered  as a rough approximation since the various $X_1$ are coarsely spaced. Nevertheless, the longitudinal turbulent transport is typically a small fraction of the lateral transport as was checked against the present two-point data as well as against the single-point transport data presented in \S 3 of \textbf{I}. 
The pressure transport, $\mathcal{T}_p^*$, data cannot be directly estimated with the present apparatus.
However, there is no \emph{a priori} reason to consider it negligible and therefore it is retained in \eqref{eq:KHM} as an unknown. 
Nevertheless, the contribution from $\mathcal{T}_p^*$ can be inferred indirectly from the deviations of the measured terms' balance via \eqref{eq:KHM}. 

\item $4\mathcal{D}^*_{\nu} \approx 2\nu/r^2\,\partial/\partial r \left(r^2\,\partial/\partial r \left(\overline{\delta q^2}\right)\right)$, i.e. only the radial component of the Laplacian is computed. 
Note that the integrals of the polar, $ \mathcal{D}^*_{\nu,\theta} $, and azimuthal, $\mathcal{D}^*_{\nu,\phi}$, components of the Laplacian over a spherical shell are identically zero,  $\oiint_{|\mathbf{r}|=r} \mathcal{D}^*_{\nu,\theta} \,dS=\oiint_{|\mathbf{r}|=r} \mathcal{D}^*_{\nu,\phi} \,dS=0$ and therefore these terms represent  the viscous diffusion across the different orientations $\mathbf{r}/r$. 
Only spherical shell averages (effectively circumferential averages) are discussed below and therefore the polar and azimuthal components are not computed.

\item $4\mathcal{D}^*_{X,\nu} \approx \nu/2\,\partial^2/\partial X_1^2\left( \overline{\delta q^2}\right)+\nu/2\,\partial^2/\partial X_2^2\left( \overline{\delta q^2}\right) \approx 0$, see Appendix \ref{sec:neglect}.

\item $4\varepsilon^*\approx 4\varepsilon^{\mathrm{iso,3}}$, i.e. the centreline energy dissipation estimate $\varepsilon^{\mathrm{iso,3}}$ is used as a surrogate for the average of the actual dissipation at $\mathbf{x}$ and $\mathbf{x'}$ (see \S 5 in \textbf{I} where the different dissipation estimates are discussed). 
Note that with the present data it is only possible to estimate the dissipation rate along the centerline. Nevertheless, the spanwise profiles of the (less suitable) surrogate $\varepsilon^{\mathrm{iso}}$ indicate that the departures from the centreline value are within $10\%$, see figure 4d in \textbf{I}. \\

\end{itemize}

Of particular importance to the subsequent discussions are the circumferential averages of the terms in \eqref{eq:KHM} in order to remove the dependence on orientation ($\mathbf{r}/r$) of the turbulence statistics.  The circumferential averages are expected to be good approximations to the averages over spherical shells considering the statistical axisymmetry of the turbulence with respect to the centreline. (Recall that for most of the present data $\mathbf{X}$, and therefore $r_1$, lies along the centreline. However, for the two datasets acquired off-centreline at $X_2=y=-6$mm one may expect the validity of this assumption to be more doubtful.)
The circumferential averages are obtained by integration with respect to the polar angle $\theta$ as $\int_{0}^{\pi/2}\,A^*(r,\,\theta,\,0)\,\mathrm{sin}(\theta)\, d\theta$, where the integrand $A^*$  is any one of the measured terms in \eqref{eq:KHM} (note that only one quarter of the domain is used due to the reflection symmetry of the structure functions around the $r_1$ and $r_2$ axes, the former due to stationarity and the latter by construction).
The wind tunnel measurements of \cite{Nagata2012} for the decay region in the lee of FSGs and the numerical data of \cite{Sylvain2011} for both FSGs and a RG gives substantial support to this assumption and therefore the circumferential averages are  interpreted as spherical shell averages throughout this paper.
The spherical shell averaged terms are denoted by removing the superscript asterisk.

\subsection{Anisotropy of energy transfer} \label{sec:aniso}
The anisotropy of the structure functions $\overline{\delta q^2}(r,\,\theta,\,\phi=0)$ and $\overline{\delta u_i\delta q^2}(r,\,\theta,\,\phi=0)$  is qualitatively investigated from their dependence on $\theta$. (For notational simplicity and due to the assumed axisymmetry, $\phi$ is not explicitly used as an argument henceforth.)
Note that in the present context anisotropy refers to the dependence of the terms in \eqref{eq:KHM}  on the orientation $\mathbf{r}/r$  \cite[see also][]{Moisy2011,Danaila2012} and not to the kinematic relation between the components of the structure functions parallel and perpendicular to $\mathbf{r}$ (e.g. $\overline{(\delta u_{\parallel})^2}$ versus $\overline{(\delta u_{\perp})^2}$ and $\overline{\delta u_{\parallel} (\delta u_{\parallel})^2}$ versus $\overline{\delta u_{\parallel}(\delta u_{\perp})^2}$), except when clearly indicated. 
The latter anisotropy considerations are complementary to the first but pertain, for example,  to the distribution of kinetic energy between the three orthogonal components and the inter-component energy transfer via pressure fluctuations \cite[see e.g. \citealt{SThesis}, ][and references therein]{SJ98}. 

\begin{figure}
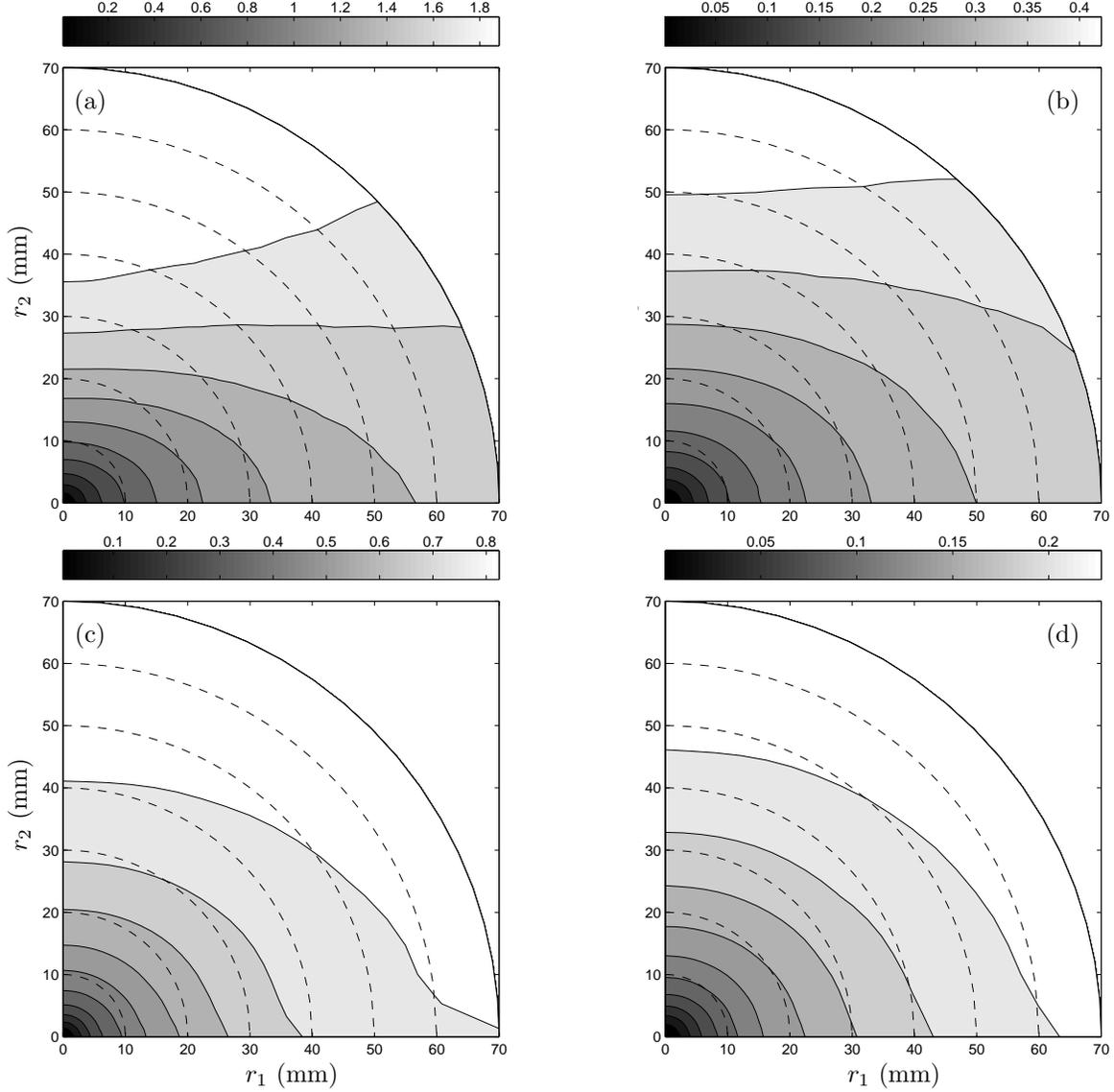

\centering
\begin{minipage}[c]{0.5\linewidth}
   \centering
   \begin{lpic}[b(-4mm)]{Figures/RG115-dq2=1250(70mm)}
   \lbl{20,110;(a)}
   \lbl[W]{3,70,90;\hspace{5mm}$r_2$ (mm)\hspace{5mm}}
  \end{lpic}
\end{minipage}%
\begin{minipage}[c]{0.5\linewidth}
   \centering 
   \begin{lpic}[b(-4mm),l(1.8mm)]{Figures/RG115-dq2=3050(70mm)}
   \lbl{110,110;(b)}
   \lbl[W]{3.7,70,90; \hspace{10mm}${}^{}$ \hspace{10mm}}
  \end{lpic}
\end{minipage}
\begin{minipage}[c]{0.5\linewidth}
   \centering
   \begin{lpic}{Figures/RG60-dq2=1250(70mm)}
   \lbl{20,110;(c)}
   \lbl[W]{3,70,90;\hspace{5mm}$r_2$ (mm)\hspace{5mm}}
   \lbl[W]{70,3;$r_1$ (mm)}
   \end{lpic}
\end{minipage}%
\begin{minipage}[c]{0.5\linewidth}
   \centering 
   \begin{lpic}[l(1.8mm)]{Figures/RG60-dq2=3050(70mm)}
   \lbl{110,110;(d)}
   \lbl[W]{3.7,70,90; \hspace{10mm}${}^{}$ \hspace{10mm}}
   \lbl[W]{70,3;$r_1$ (mm)}
   \end{lpic}
\end{minipage}
\caption[Second order structure functions, $\overline{\delta q^2}(r_1,r_2)$]{Iso-contours of the second-order structure functions, $\overline{\delta q^2}(r_1,r_2)$ (m$^2$s$^{-2}$), at (a,c) $X_1=1250$mm and (b,d) $X_1=3050$mm for (top) RG115 and (bottom) RG60 data.  $X_1=1250$mm and $X_1=3050$mm correspond to $X_1/x_{\mathrm{peak}} = 1.5$ and $X_1/x_{\mathrm{peak}} = 3.7$ for the RG115 data and to $X_1/x_{\mathrm{peak}} = 8.5$ and $X_1/x_{\mathrm{peak}} = 20.7$ for the RG60 data. The reference contour levels for isotropic turbulence are added as dashed lines.}
\label{fig:2ndStrut}
\end{figure}

The second-order structure functions $\overline{\delta q^2}(r,\,\theta)$ are presented in figures \ref{fig:2ndStrut}a-d for the furthermost upstream and downstream measurement locations and for turbulence generated by both RG115 and RG60.
Comparing the upstream data (figures \ref{fig:2ndStrut}a,c) with the downstream data (figures \ref{fig:2ndStrut}b,d) for both grids there seems to be a tendency for the contours to become increasingly circular as the turbulence decays, i.e. for the energy distribution to become increasingly isotropic.
Furthermore, comparing the RG115 with the RG60 data (figures \ref{fig:2ndStrut}a,b and \ref{fig:2ndStrut}c,d, respectively) it can be seen that the RG115 data, which are acquired closer to the grid in terms of $x_{\mathrm{peak}}$ multiples, is less isotropic.
Both these observations corroborate a tendency for the kinetic energy to become uniformly distributed over spherical shells for larger $x/x_{\mathrm{peak}}$. 
Nevertheless, for all cases the small scales seem to remain anisotropic by velocity derivative measures (see \S 5 and tables 4  and 5 of \textbf{I}).

\begin{figure}
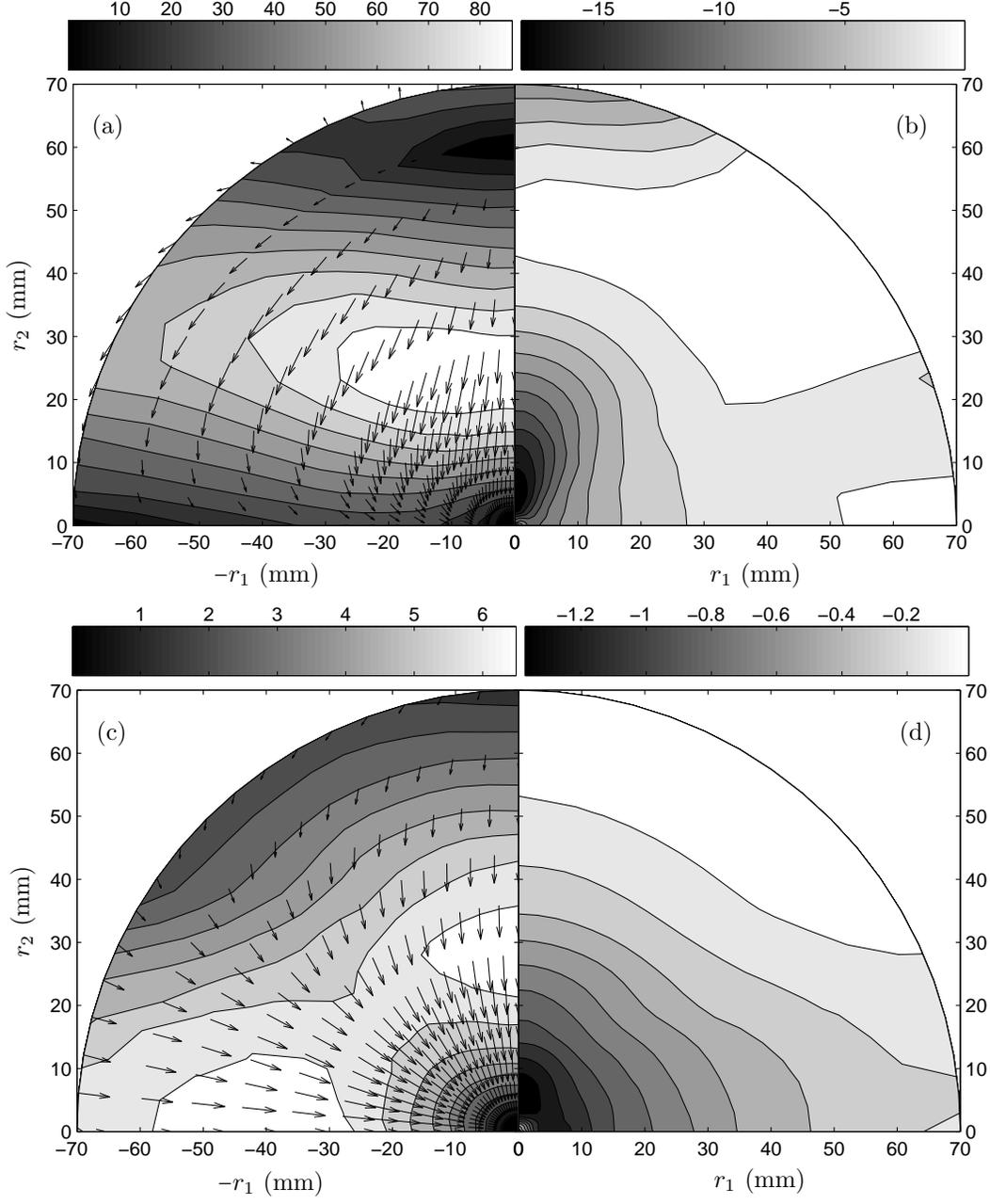

\centering
\begin{lpic}{Figures/RG115-ModAndDiv_duidq2=1250(140mm)}
   \lbl{20,86;(a)}
   \lbl{168,86;(b)}
   \lbl[W]{3.6,54,90;\hspace{10mm}$r_2$ (mm)\hspace{10mm}}
   \lbl[W]{50,3.5;\hspace{10mm}$-r_1$ (mm)\hspace{10mm}}
   \lbl[W]{140,3.5;\hspace{10mm}$r_1$ (mm)\hspace{10mm}}
\end{lpic}
\begin{lpic}{Figures/RG115-ModAndDiv_duidq2=3050(140mm)}
   \lbl{20,86;(c)}
   \lbl{168,86;(d)}
   \lbl[W]{3.6,54,90;\hspace{10mm}$r_2$ (mm)\hspace{10mm}}
   \lbl[W]{50,3;\hspace{10mm}$-r_1$ (mm)\hspace{10mm}}
   \lbl[W]{140,3.5;\hspace{10mm}$r_1$ (mm)\hspace{10mm}}
\end{lpic}
\caption{(a,c) Third-order structure function vectors, $\overline{\delta u_i \delta q^2}$ and iso-contours of their magnitude, $|\overline{\delta u_i \delta q^2}| \,(\times 10^{-3}\,\, \mathrm{m^3s^{-3}})$. (b,d) Iso-contours of the radial contribution of the divergence of $\overline{\delta u_i \delta q^2}$, $\Pi^*_r \,(\mathrm{m^2s^{-3}})$. 
(top) $X_1=1250$mm and (bottom) $X_1=3050$mm. $X_1=1250$mm and $X_1=3050$mm correspond to $X_1/x_{\mathrm{peak}} = 1.5$ and $X_1/x_{\mathrm{peak}} = 3.7$.
Data are acquired in the lee of RG115.}
\label{fig:3rdStrutRG115}
\end{figure}

\begin{figure}
\centering
\begin{lpic}{Figures/RG60-ModAndDiv_duidq2=1250(140mm)}
   \lbl{20,86;(a)}
   \lbl{168,86;(b)}
   \lbl[W]{3.6,54,90;\hspace{10mm}$r_2$ (mm)\hspace{10mm}}
   \lbl[W]{50,3.5;\hspace{10mm}$-r_1$ (mm)\hspace{10mm}}
   \lbl[W]{140,3.5;\hspace{10mm}$r_1$ (mm)\hspace{10mm}}
\end{lpic}
\begin{lpic}{Figures/RG60-ModAndDiv_duidq2=3050(140mm)}
   \lbl{20,86;(c)}
   \lbl{168,86;(d)}
   \lbl[W]{3.6,54,90;\hspace{10mm}$r_2$ (mm)\hspace{10mm}}
   \lbl[W]{50,3;\hspace{10mm}$-r_1$ (mm)\hspace{10mm}}
   \lbl[W]{140,3.5;\hspace{10mm}$r_1$ (mm)\hspace{10mm}}
\end{lpic}
\caption{(a,c) Third-order structure function vectors, $\overline{\delta u_i \delta q^2}$ and iso-contours of their magnitude, $|\overline{\delta u_i \delta q^2}| \,(\times 10^{-3}\,\, \mathrm{m^3s^{-3}})$. (b,d) Iso-contours of the radial contribution of the divergence of $\overline{\delta u_i \delta q^2}$, $\Pi^{*}_r \,(\mathrm{m^2s^{-3}})$. 
(top) $X_1=1250$mm and (bottom) $X_1=3050$mm.
Data are acquired in the lee of RG60. $X_1=1250$mm and $X_1=3050$mm correspond to $X_1/x_{\mathrm{peak}} = 8.5$ and $X_1/x_{\mathrm{peak}} = 20.7$.}
\label{fig:3rdStrutRG60}
\end{figure}

Turning to the third-order structure function vectors $\overline{\delta u_i\delta q^2}(r,\,\theta)$, a similar tendency to isotropy is observed (figures \ref{fig:3rdStrutRG115}a,c and \ref{fig:3rdStrutRG60}a,c). The third-order structure function vectors, which for the RG115 data at $x = 1.5x_{\mathrm{peak}}$ are nearly aligned with the tangential direction (figure \ref{fig:3rdStrutRG115}a), progressively align with the radial direction and for the RG60 data at $x = 21x_{\mathrm{peak}}$ (figure \ref{fig:3rdStrutRG60}c) they are indeed nearly so. 
Note that the divergence of $\overline{\delta u_i\delta q^2}$ (i.e. $\Pi^*$) has a radial and a polar contributions (the azimuthal contribution is taken to be zero due to the assumed axisymmetry). 
As discussed in \S \ref{sec:KHM}, the radial contribution $\Pi_r^*$ (which we plot in figures \ref{fig:3rdStrutRG115}b,d and \ref{fig:3rdStrutRG60}b,d) relates to the interscale energy transfer, whereas the polar contribution $\Pi_{\theta}^*$ accounts for the redistribution of energy within a spherical shell.
The above mentioned tendency to isotropy as the flow decays is very likely linked to the redistribution of energy via $\Pi_{\theta}^*$. 

\section{The role of turbulence production and transport} \label{sec:inhomo}
The effect of transport and production in the single-point kinetic energy balance was investigated in \textbf{I} where it was found that, for the assessed region of the RG115-generated turbulence, both contributions are non-negligible by comparison with the energy dissipation. 
This region of the RG115-generated turbulence was also compared with an equivalent region of turbulence generated by FSGs and considerable differences were found in the downstream evolution (and transverse profiles) of  transport and production relative to the dissipation. 
Nevertheless, the two different turbulent flows were found to have the same non-classical dissipation behaviour and, consequently, the differences in the production and transport reinforced the view that this non-classical  behaviour is present irrespective of the details of the inhomogeneity of the turbulent flow. 
Indeed, the turbulent transport and production are expected to  be large-scale phenomena that play no direct role in the scale-by-scale energy transfer mechanisms, even at these Reynolds numbers ($Re_{\lambda}=\mathcal{O}(100)$). 
Here, we present data  which allow a precise quantification of the effect of production and transport on the scale-by-scale energy budget \eqref{eq:KHM}. 

One may average \eqref{eq:KHM} over spherical shells to eliminate the dependence of each term on the  orientation $\mathbf{r}/r$ yielding the average contribution of each scale to the balance. 
Retaining all terms except for the linear transfer of energy caused by mean velocity gradients and the scale-by-scale transport by viscous diffusion which were shown to be
negligible (see Appendix \ref{sec:neglect}), the spherical averaged scale-by-scale energy balance reads, 
\begin{equation}
\mathcal{A} +  \Pi -  \mathcal{P}  - \mathcal{T} - \mathcal{T}_p =   \mathcal{D}_{\nu}  -  \varepsilon,
\label{eq:KHMSimplified}
\end{equation}
where the $\mathcal{T}$ represents the measured component of turbulent transport and  $\mathcal{T}_p$ represents the unknown contribution from the pressure transport.

\begin{figure}
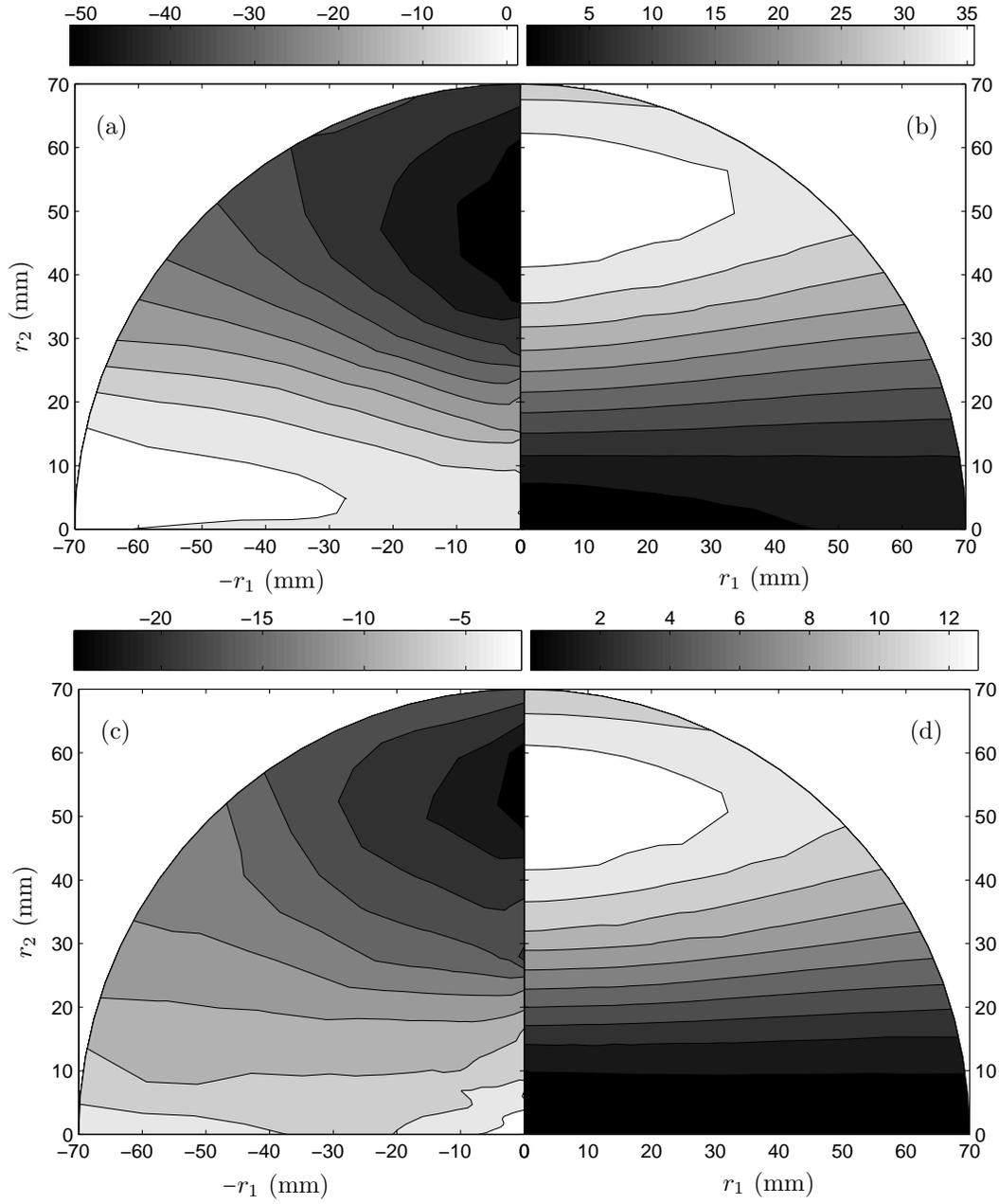

\centering
\begin{lpic}{Figures/TranspAndProd=1250(140mm)}
   \lbl{20,86;(a)}
   \lbl{168,86;(b)}
   \lbl[W]{3.5,54,90;\hspace{10mm}$r_2$ (mm)\hspace{10mm}}
   \lbl[W]{50,3;\hspace{10mm}$-r_1$ (mm)\hspace{10mm}}
   \lbl[W]{140,3.5;\hspace{10mm}$r_1$ (mm)\hspace{10mm}}
\end{lpic}
\begin{lpic}{Figures/TranspAndProd=2150(140mm)}
   \lbl{20,86;(c)}
   \lbl{168,86;(d)}
   \lbl[W]{3.5,54,90;\hspace{10mm}$r_2$ (mm)\hspace{10mm}}
   \lbl[W]{50,3;\hspace{10mm}$-r_1$ (mm)\hspace{10mm}}
   \lbl[W]{140,3.5;\hspace{10mm}$r_1$ (mm)\hspace{10mm}}
\end{lpic}
\caption{Normalised turbulence (a,c) transport, $\mathcal{T}^*/\varepsilon\,(\%)$ and (b,d) production, $\mathcal{P}^*/\varepsilon\,(\%)$ versus $(r_x,\,r_y)$ at $x/x_{\mathrm{peak}} = 1.5$ (top) and $x/x_{\mathrm{peak}} = 2.6$ (bottom) in the lee of RG115.}
\label{fig:inhomo}
\end{figure}

\begin{figure}
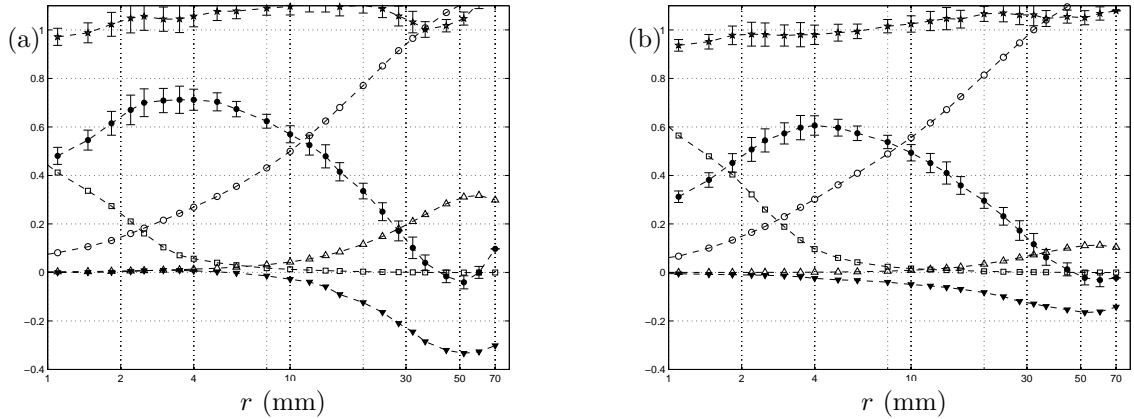

\centering
\begin{minipage}[c]{0.5\linewidth}
   \centering
   \begin{lpic}{Figures/RG115-KHMAzAve=1250(65mm)}
   \lbl{0,125;(a)}
   \lbl[W]{90,1;$r$ (mm)}
  \end{lpic}
\end{minipage}%
\begin{minipage}[c]{0.5\linewidth}
   \centering 
   \begin{lpic}{Figures/RG115-KHMAzAve=2150(65mm)}
   \lbl{0,125;(b)}
   \lbl[W]{90,1;$r$ (mm)}
   \end{lpic}
\end{minipage}
\caption{Spherically averaged (\protect\raisebox{-0.5ex}{\FilledSmallCircle}) $-\Pi/\varepsilon$, (\protect\raisebox{-0.5ex}{\SmallCircle})  $-\mathcal{A}/\varepsilon$,  (\protect\raisebox{-0.5ex}{\SmallSquare}) $\mathcal{D}_{\nu}/\varepsilon$, ($\medtriangleup$) $\mathcal{P}/\varepsilon$, ($\filledmedtriangledown$) $\mathcal{T}/\varepsilon$ and (\small{\ding{73}}\normalsize) $(\mathcal{D}_{\nu}  + \mathcal{P} + \mathcal{T} - \Pi - \mathcal{A})/\varepsilon$ at (a) $x/x_{\mathrm{peak}} = 1.5$ and (b) $x/x_{\mathrm{peak}} = 2.6$ for RG115-generated turbulence. The size of the error bars in the energy transfer data are discussed in \S \ref{sec:conv}. Error bars of equal size are added to the data points representing $(\mathcal{D}_{\nu}  + \mathcal{P} + \mathcal{T} - \Pi - \mathcal{A})$, however, this underestimates the error margins as it does not take into account the uncertainty associated with the estimates of the other terms.}
\label{fig:KHMwInhomo}
\end{figure}

Turning to the data, the iso-contour maps of the transport and production terms normalised by the dissipation, $\varepsilon$, indicate that most of the transport and production occur at $r \gtrapprox L_{11}^{(1)} \approx 30$mm and $\theta \approx \pi/2$ (figures \ref{fig:inhomo}a,c and \ref{fig:inhomo}b,d). 
At smaller values of $r$ both $\mathcal{T}$ and $\mathcal{P}$ are less than about $15\%$ of $\varepsilon$. 
Note that the production for large $r$ is much smaller for $\theta \approx 0$ than for $\theta \approx \pi/2$ because $\partial U/\partial y$ tends to zero at the centreline and the remaining production term, $2\overline{(\delta u)^2} \partial U/\partial x$, is small by comparison.  
Similarly, the transport for large $r$ and $\theta \approx 0$ is also smaller because the lateral transport overwhelms the longitudinal transport.

The spherical averaged contribution of these terms to the balance \eqref{eq:KHMSimplified} are plotted together with the spherical shell averaged advection, energy transfer and viscous diffusion in figures \ref{fig:KHMwInhomo}a,b.
Note that the dissipation estimates are compensated for the resolution of the sensor, see \S 2.2.1 in \textbf{I}.  The finite resolution of the sensor also biases $\mathcal{D}_{\nu}$ since  $\lim_{r\rightarrow 0}\mathcal{D}_{\nu}(r)=\varepsilon$. A rough compensation for this bias is applied by multiplying $\mathcal{D}_{\nu}$ with the ratio between the corrected and the measured $\varepsilon$.)

The radial distributions of the advection, energy transfer and viscous diffusion are similar to those found in the literature for data at comparable Reynolds numbers \cite[see e.g.][]{Antonia2006}. 
From the data it is clear that our turbulent transport and production terms are significant for scales of the order of the integral-length scale but become negligible at scales smaller than $r \approx 10\mathrm{mm}\approx L_{11}^{(1)}/3$  and therefore cannot tamper with the scale-by-scale energy transfer around its maximum ($r \approx 4\mathrm{mm}\approx L_{11}^{(1)}/8$ for the present data).
This provides  quantitative evidence that the influence of the turbulence production and transport on the energy transfer mechanisms is negligible.

Note that in figures \ref{fig:KHMwInhomo}a,b the balance of the measured terms is also presented. 
By virtue of \eqref{eq:KHMSimplified}, the scale-by-scale advection, energy transfer, production, transport and viscous diffusion should balance the dissipation plus the unknown contribution from scale-by-scale pressure transport, $\mathcal{T}_p$. 
Even though $\mathcal{T}_p$ is not accounted for, it can be seen that there is a reasonable balance between the measured terms, at least within the expected uncertainty of the data. 
Note that the error bars added to the balance ($\mathcal{D}_{\nu}  + \mathcal{P} + \mathcal{T} - \Pi - \mathcal{A}$, see figures \ref{fig:KHMwInhomo}a,b) underestimate the overall uncertainty of the data since they do not take into account  uncertainties associated with the measurements of the advection, transport and production terms and possible departures from the assumptions used to compute the terms in \eqref{eq:KHM}, see \S \ref{sec:KHMcomputed}.

\section{Advection, energy transfer and dissipation scalings} \label{sec:5}
We now investigate how the stark differences in the way the energy dissipation scales in the two decay regions identified in \cite{VV2012} downstream of a turbulence-generating grid relate to the advection, energy transfer and viscous diffusion during decay (the remaining terms in  \eqref{eq:KHMSimplified} are negligible at small enough values of $r$ as shown in figure \ref{fig:KHMwInhomo}).

\begin{figure}
\centering 
\begin{lpic}[b(0mm)]{Figures/Paper_RG60KHMnormyEpsnormxNone(130mm)}
\lbl{5,130;(a)}
\lbl{70,82;$-\Pi/\varepsilon$}
\lbl{75,28;$\mathcal{D}_{\nu}/\varepsilon$}
\lbl{155,105;$-\mathcal{A}/\varepsilon$}
\lbl{45,130;$(\mathcal{D}_{\nu} - \Pi - \mathcal{A})/\varepsilon$}
\lbl[W]{4,80,90;\hspace{10mm}($\mathcal{A}$, $\Pi$, $\mathcal{D}_{\nu}$)/$\varepsilon$\hspace{10mm}}
\lbl[W]{90,4;\hspace{10mm}$r$ (mm)\hspace{10mm}}
\end{lpic}
\begin{lpic}[b(0mm)]{Figures/Paper_RG60KHMnormyEpsnormxLamb(130mm)}
\lbl{5,130;(b)}
\lbl{74,82;$-\Pi/\varepsilon$}
\lbl{75,25;$\mathcal{D}_{\nu}/\varepsilon$}
\lbl{140,105;$-\mathcal{A}/\varepsilon$}
\lbl{45,130;$(\mathcal{D}_{\nu} - \Pi - \mathcal{A})/\varepsilon$}
\lbl[W]{4,80,90;\hspace{10mm}($\mathcal{A}$, $\Pi$, $\mathcal{D}_{\nu}$)/$\varepsilon$\hspace{10mm}}
\lbl[W]{90,4;\hspace{10mm}$r/\lambda$ \hspace{10mm}}
\end{lpic}
\caption[$\Pi/\varepsilon$, $\mathcal{A}/\varepsilon$ and $\mathcal{D}_{\nu}/\varepsilon$ throughout the decay of RG60-generated turbulence]{Normalised, spherical shell averaged scale-by-scale energy transfer ($-\Pi/\varepsilon$), advection ($-\mathcal{A}/\varepsilon$) and viscous diffusion ($\mathcal{D}_{\nu}/\varepsilon$) versus (a) $r$ and (b) $r/\lambda$, during the decay of turbulence generated by  RG60 at  (\protect\raisebox{-0.5ex}{\SmallCircle}) $x/x_{\mathrm{peak}} = 8.5$, (\protect\raisebox{-0.5ex}{\SmallSquare}) $x/x_{\mathrm{peak}} = 11.5$, ($\medtriangleright$) $x/x_{\mathrm{peak}} = 16.6$, (\protect\raisebox{-0.5ex}{\Diamondshape}) $x/x_{\mathrm{peak}} = 17.6$  and  (\ding{73}) $x/x_{\mathrm{peak}} = 21$. The 95\% confidence intervals of the normalised $\Pi$ (see \S \ref{sec:conv}) are added to the furthermost up- and downstream locations.}
\label{fig:DownEvoRG60}
\end{figure}

\begin{figure}
\centering
\begin{lpic}[b(0mm)]{Figures/Paper_RG115KHMnormyEpsnormxNone(130mm)}
\lbl{5,130;(a)}
\lbl{80,92;$-\Pi/\varepsilon$}
\lbl{75,28;$\mathcal{D}_{\nu}/\varepsilon$}
\lbl{140,105;$-\mathcal{A}/\varepsilon$}
\lbl{70,132;$(\mathcal{D}_{\nu} - \Pi - \mathcal{A})/\varepsilon$}
\lbl[W]{4,80,90;\hspace{10mm}($\mathcal{A}$, $\Pi$, $\mathcal{D}_{\nu}$)/$\varepsilon$\hspace{10mm}}
\lbl[W]{90,4;\hspace{10mm}$r$ (mm)\hspace{10mm}}
\end{lpic}
\begin{lpic}[b(0mm)]{Figures/Paper_RG115KHMnormyEpsnormxLamb(130mm)}
\lbl{5,130;(b)}
\lbl{80,95;$-\Pi/\varepsilon$}
\lbl{75,25;$\mathcal{D}_{\nu}/\varepsilon$}
\lbl{140,105;$-\mathcal{A}/\varepsilon$}
\lbl{70,132;$(\mathcal{D}_{\nu} - \Pi - \mathcal{A})/\varepsilon$}
\lbl[W]{4,80,90;\hspace{10mm}($\mathcal{A}$, $\Pi$, $\mathcal{D}_{\nu}$)/$\varepsilon$\hspace{10mm}}
\lbl[W]{90,4;\hspace{10mm}$r/\lambda$ \hspace{10mm}}
\end{lpic}
\caption[$\Pi/\varepsilon$, $\mathcal{A}/\varepsilon$ and $\mathcal{D}_{\nu}/\varepsilon$ throughout the decay of RG115-generated turbulence]{Normalised, spherical shell averaged scale-by-scale energy transfer ($-\Pi/\varepsilon$), advection ($-\mathcal{A}/\varepsilon$) and viscous diffusion ($\mathcal{D}_{\nu}/\varepsilon$) versus (a) $r$ and (b) $r/\lambda$, during the decay of turbulence generated by  RG115 at  (\protect\raisebox{-0.5ex}{\SmallCircle}) $x/x_{\mathrm{peak}} = 1.5$, (\protect\raisebox{-0.5ex}{\SmallSquare}) $x/x_{\mathrm{peak}} = 2.0$, ($\medtriangleright$) $x/x_{\mathrm{peak}} = 2.6$, (\protect\raisebox{-0.5ex}{\Diamondshape}) $x/x_{\mathrm{peak}} = 3.1$ and  (\ding{73}) $x/x_{\mathrm{peak}} = 3.7$. The 95\% confidence intervals of the normalised $\Pi$ (see \S \ref{sec:conv}) are added to the furthermost up- and downstream locations.}%
\label{fig:DownEvoRG115}
\end{figure}

\begin{figure}
\centering
\begin{lpic}[b(0mm)]{Figures/Paper_RG115KHMnormyq3normxLamb(130mm)}
\lbl{65,75;$-\Pi\,L_{11}^{(1)}/(\overline{u^2})^{3/2}$}
\lbl{85,12;$\mathcal{D}_{\nu}\,L_{11}^{(1)}/(\overline{u^2})^{3/2}$}
\lbl{147,86;$-\mathcal{A}\,L_{11}^{(1)}/(\overline{u^2})^{3/2}$}
\lbl[W]{-5,80,90;\hspace{15mm}($\mathcal{A}$, $\Pi$, $\mathcal{D}_{\nu}$)$\,L_{11}^{(1)}/(\overline{u^2})^{3/2}$\hspace{15mm}}
\lbl[W]{90,-4;\hspace{10mm}$r/\lambda$ \hspace{10mm}}
\end{lpic}
\vspace{10mm}
\caption[$\Pi \,L_{11}^{(1)}/(\overline{q^2})^{3/2}$, $\mathcal{A} \,L_{11}^{(1)}/(\overline{u^2})^{3/2}$ and $\mathcal{D}_{\nu}\,L_{11}^{(1)}/(\overline{u^2})^{3/2}$ throughout the decay of RG115-generated turbulence]{Normalised, spherical shell averaged scale-by-scale energy transfer ($-\Pi \,L_{11}^{(1)}/(\overline{u^2})^{3/2}$), advection ($-\mathcal{A}\,L_{11}^{(1)}/(\overline{u^2})^{3/2}$) and viscous diffusion ($\mathcal{D}_{\nu} \,L_{11}^{(1)}/(\overline{u^2})^{3/2}$) versus $r/\lambda$, during the decay of turbulence generated by  RG115 at  (\protect\raisebox{-0.5ex}{\SmallCircle}) $x/x_{\mathrm{peak}} = 1.5$, (\protect\raisebox{-0.5ex}{\SmallSquare}) $x/x_{\mathrm{peak}} = 2.0$, ($\medtriangleright$) $x/x_{\mathrm{peak}} = 2.6$, (\protect\raisebox{-0.5ex}{\Diamondshape}) $x/x_{\mathrm{peak}} = 3.1$ and  (\ding{73}) $x/x_{\mathrm{peak}} = 3.7$. The 95\% confidence intervals of the normalised $\Pi$ (see \S \ref{sec:conv}) are added to the furthermost up- and downstream locations.}
\label{fig:DownEvoRG115b}
\end{figure}

Starting with the RG60 data, the evolution in the further downstream decay region between  $x= 8.5x_{\mathrm{peak}}$ and $x= 21x_{\mathrm{peak}}$ of the scale-by-scale viscous diffusion, energy transport and advection normalised by the dissipation are shown in figure \ref{fig:DownEvoRG60}a. 
As the turbulence decays these terms seem to move to the right reflecting the increase in the turbulent scales. 
Normalising the abscissae by $\lambda$ seems to account for much of the spread (figure \ref{fig:DownEvoRG60}b).  
The scaling of the abscissae is, however, secondary to the main discussion here which pertains to the relative magnitude of the advection, the energy transfer, the viscous diffusion and the dissipation. 
Note that the viscous diffusion is very small compared to the dissipation at scales $r\ge \lambda$ for both our RG60 data (see figure \ref{fig:DownEvoRG60}b) and our RG115 data (figure \ref{fig:DownEvoRG115}b), in agreement with a mathematical proof of this fact which we give in Appendix \ref{sec:appendix}.
Of particular importance in figure \ref{fig:DownEvoRG60}b is the observation that the maximum absolute value of the energy transfer $\Pi |_{\mathrm{max}}$ is roughly a constant fraction of the dissipation throughout the downstream extent of the data corresponding to a range of local Reynolds numbers $Re_{\lambda}$ between 100 and 80 ($-\Pi |_{\mathrm{max}}\approx 0.55 \varepsilon$ with the peak located at $r\approx\lambda$, see figure \ref{fig:DownEvoRG60}b).

\begin{figure}
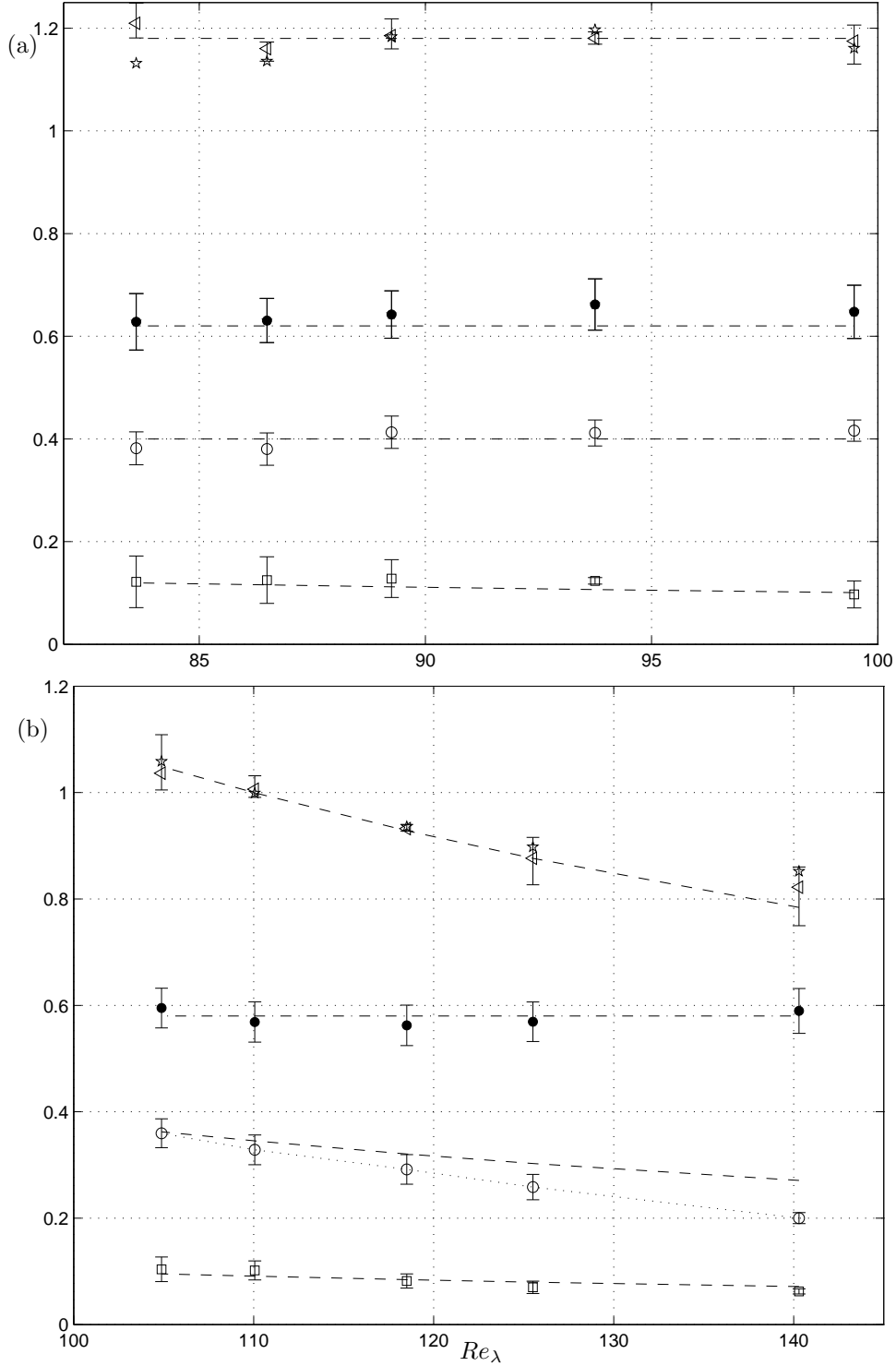

\centering
\begin{lpic}[]{Figures/Paper_RG60TermsNormTurnover(125mm)}
\lbl{-2,120;(a)}
\end{lpic}
\begin{lpic}[]{Figures/Paper_RG115TermsNormTurnover(122mm)}
\lbl{-2,120;(b)}
\lbl[W]{86,-1;$Re_{\lambda}$}
\end{lpic}
\vspace{5mm}
\caption[$C_{\varepsilon}^{1(1)}$,   $C_{\Pi}^{1(1)}$,  $-\mathcal{A}|_{\mathrm{max}(\Pi)}L_{11}^{(1)}/(\overline{u^2})^{3/2}$ and $\mathcal{D}_{\nu}|_{\mathrm{max}(\Pi)}L_{11}^{(1)}/(\overline{u^2})^{3/2}$ versus $Re_{\lambda}$ during the decay of turbulence generated by RG60 and RG115]{Normalised energy dissipation, maximum scale-by-scale energy transfer and scale-by-scale advection and viscous diffusion at the maximum energy transfer versus $Re_{\lambda}$ during the decay of turbulence generated by (a) RG60 and (b) RG115;
 ($\medtriangleleft$) $C_{\varepsilon}^{1(1)}$,   (\protect\raisebox{-0.5ex}{\FilledSmallCircle}) $C_{\Pi}^{1(1)} \equiv  -\Pi |_{\mathrm{max}}L_{11}^{(1)}/(\overline{u^2})^{3/2}$, (\protect\raisebox{-0.5ex}{\SmallCircle}) $-\mathcal{A}|_{\mathrm{max}(\Pi)}L_{11}^{(1)}/(\overline{u^2})^{3/2}$,  (\protect\raisebox{-0.5ex}{\SmallSquare}) $\mathcal{D}_{\nu}|_{\mathrm{max}(\Pi)}L_{11}^{(1)}/(\overline{u^2})^{3/2}$ and (\small{\ding{73}}\normalsize) $(\mathcal{D}_{\nu}|_{\mathrm{max}(\Pi)}\, - \,\Pi |_{\mathrm{max}}\,-\,\mathcal{A}|_{\mathrm{max}(\Pi)})L_{11}^{(1)}/(\overline{u^2})^{3/2}$. Dash and dash-dot lines follow $\sim {Re}_{\lambda}^{-1}$ and $\sim {Re}_{\lambda}^{0}$, respectively.}
\label{fig:PeakValues}
\end{figure}

In fact, taking the numerical values of $\Pi |_{\mathrm{max}}$ and the numerical values of the advection at the separation $r^*$ where $\Pi(r^*) = \Pi |_{\mathrm{max}}$ and normalising the data with $(\overline{u^2})^{3/2}/L_{11}^{(1)}$ it is clear from figure \ref{fig:PeakValues}a  that $-\mathcal{A}|_{\mathrm{max}(\Pi)}L_{11}^{(1)}/(\overline{u^2})^{3/2}\sim C_{\Pi}^{1(1)}\sim C_{\varepsilon}^{1(1)}\approx \mathrm{constant}$  (where $C_{\varepsilon}^{1(1)}\equiv \varepsilon L_{11}^{(1)}/(\overline{u^2})^{3/2}$, $C_{\Pi}^{1(1)}\equiv- \Pi |_{\mathrm{max}}L_{11}^{(1)}/(\overline{u^2})^{3/2}$ and $ L_{11}^{(1)}$ is the usual longitudinal integral length-scale - for further details refer to \textbf{I}).
The viscous diffusion term, $\mathcal{D}_{\nu}|_{\mathrm{max}(\Pi)}$ is smaller than any of the other terms at this moderate $Re_{\lambda}$ ($<10\%$ of the dissipation) and it is difficult to discern whether $\mathcal{D}_{\nu}|_{\mathrm{max}(\Pi)}$ is constant or decreases with increasing $Re_{\lambda}$ as one might expect. 

Turning to the RG115 data presented in figure \ref{fig:DownEvoRG115}a two outstanding differences in the downstream evolution of these quantities can be registered: (i) the peak value of the energy transfer does not scale with the dissipation and (ii) the curves representing the advection term are moving from right to left, in the opposite direction than was the case for the RG60 data (figure \ref{fig:DownEvoRG60}a).
Normalising the abscissae with $\lambda$ takes into account most of the spread in the viscous diffusion term but now augments the spread of the advection term (see figure \ref{fig:DownEvoRG115}b and compare with  figure \ref{fig:DownEvoRG60}b).
(Note that for the RG115 data in this region, $L_{11}^{(1)}\sim \lambda$ as shown in \cite{VV2012}, hence the normalisation of the abscissae with $L_{11}^{(1)}$ would yield an identical horizontal collapse as that presented in figure \ref{fig:DownEvoRG115}b). 
Concerning the scaling of the ordinates, it should be noted that, if instead of $\varepsilon$ one chooses to normalise the ordinates by $(\overline{u^2})^{3/2}/L_{11}^{(1)}$ (figure \ref{fig:DownEvoRG115b}) the vertical spread of the energy transfer data is much reduced, but the spread of the advection is further augmented (as is the spread of the viscous diffusion term, since in the limit $r\rightarrow 0$ this term is equal to the dissipation and, as shown in \cite{VV2012}, $\varepsilon$ does not scale with $(\overline{u^{2}})^{3/2}/L_{11}^{(1)}$ in this region).  

The procedure of normalising $\varepsilon$, $\Pi |_{\mathrm{max}}$, $\mathcal{A}|_{\mathrm{max}(\Pi)}$ and $\mathcal{D}_{\nu}|_{\mathrm{max}(\Pi)}$ with   $(\overline{u^2})^{3/2}/L_{11}^{(1)}$ is repeated and the data are plotted in figure \ref{fig:PeakValues}b against $Re_{\lambda}$. 
Even though the dissipation follows $C_{\varepsilon}^{1(1)} = f(Re_M)/Re_{\lambda}$ in this region it is clear that the behaviour of $C_{\Pi}^{1(1)}$ is strikingly different. 
In fact, $C_{\Pi}^{1(1)}$ is approximately constant and with the same numerical value ($C_{\Pi}^{1(1)}\approx 0.6$) as the one that we find for the RG60 data in the further downstream region (in multiples of $x_{\mathrm{peak}}$) where $C_{\varepsilon}^{1(1)}$ is approximately constant.
Note also that the normalised advection term grows faster than $Re_{\lambda}^{-1}$ with decreasing $Re_{\lambda}$ and therefore adapts to cover most of the growing difference between the constant  $C_{\Pi}^{1(1)}$ and the increasing $C_{\varepsilon}^{1(1)}$ as the flow decays and $Re_{\lambda}$ decreases. 
The viscous diffusion term $\mathcal{D}_{\nu}|_{\mathrm{max}(\Pi)}$ is also small for the present data, similar to what is  found for the RG60 data.

\subsection{Discussion}\label{sec:FRN}

The present work is concerned with the validity of the energy transfer/dissipation balance over a range of length-scales $r$, i.e.
\begin{equation}
 \Pi(\mathbf{X},r) =  \Pi |_{\mathrm{max}}(\mathbf{X}) =-\varepsilon(\mathbf{X}), 
\label{eq:PiEps}
\end{equation}
where instead of using (local) isotropy, $\Pi^*(\mathbf{X},\mathbf{r})$ is averaged over spherical shells \cite[]{NT99}. 
Note that we have approximated our flow as being locally homogeneous to remove the dependence of the right-hand-side on the separation $r$. We do so based on our RG115 data where $\mathrm{max}(\Pi)$ is located at $r\simeq 5$mm, \emph{cf.} figure  \ref{fig:DownEvoRG115}a, corresponding to $y/M \simeq \pm 0.02$; for such close locations no appreciable changes in the dissipation rate can be observed, see figure 4d in \textbf{I}.
Finally, note also that $\Pi |_{\mathrm{max}}$ in physical space is equal to its wavenumber space counterpart, see appendix \ref{sec:appendixB}.%

It is clear from the outset that the Reynolds numbers of the present data are insufficiently high to allow verification of  \eqref{eq:PiEps} over a range of length-scales $r$. 
What the present data do allow us to report for the first time, however, is that  $\Pi |_{\mathrm{max}}\sim(\overline{u^2})^{3/2}/L_{11}^{(1)}$ \cite[see][]{McComb2010} both
in far downstream equilibrium turbulence where $C_{\varepsilon}^{1(1)} \approx \mathrm{constant}$ and in non-equilibrium turbulence where $C_{\varepsilon}^{1(1)} = f(Re_{M})/Re_{\lambda}$ and $Re_{\lambda}$ is higher (see figure \ref{fig:PeakValues}a,b). 
In this non-equilibrium region our data also demonstrate the growing importance of the small-scale advection with increasing streamwise distance from the grid. 
This increasing importance is directly linked to the growing imbalance between $\Pi |_{\mathrm{max}}$ and $\varepsilon$ (see figure \ref{fig:PeakValues}b and recall that $Re_{\lambda}$ decreases with increasing streamwise distance in the decay region downstream of the turbulence-generating grid).
Note also that the increasing imbalance varies too steeply with $Re_{\lambda}$ compared to the FRN effects discussed by \cite{Qian99,Moisy1999,Lundgren2002,Lundgren2003,Gagne2004,Antonia2006,Cambon2012}. 
In fact, for the Reynolds number range straddled in the present experiments the expected effect of the FRN should be constant throughout the decay as is the case for the equilibrium data in the lee of RG60.

Even though the constancy of $C_{\varepsilon}^{1(1)}$ is usually expected as a high Reynolds number asymptotic, the equilibrium $C_{\varepsilon}^{1(1)}$ constancy appears at distances much further downstream in our experiments where the local Reynolds number has in fact further decayed (though, clearly, not enough for $C_{\varepsilon}^{1(1)}$ not to be constant as a result of the local Reynolds number being too low). 
The constancy of $C_{\varepsilon}^{1(1)}$ in this far downstream equilibrium region appears in our RG60 experiments with a rate of change of the local Reynolds number which
is enough for $C_{\varepsilon}^{1(1)}$ to vary in proportion to $1/Re_{\lambda}$ in the non-equilibrium region (see and compare figures \ref{fig:PeakValues}a and \ref{fig:PeakValues}b). 
Note the high value of the constant $C_{\varepsilon}^{1(1)}$ in the RG60 equilibrium decay region experiment (figure \ref{fig:PeakValues}a), high by comparison to values of this constant recorded for forced statistically stationary turbulence and in agreement with
time-lag non-equilibrium arguments \cite[]{Bos2007}. 
What we call the far downstream equilibrium decay region may in fact be a time-lag
non-equilibrium region in the terms of \cite{Bos2007}. 
Note that this is a region where there is a "balance" between the scalings of
$\Pi |_{\mathrm{max}}$ and $\varepsilon$ (i.e. they both scale as $(\overline{u^{2}})^{3/2}/L_{11}^{(1)}$) whereas such a scaling balance is absent in what we term the non-equilibrium decay region.

\section{Conclusion}\label{sec:conclusion}
An experimental investigation of the downstream evolution of the scale-by-scale energy transfer budget for both equilibrium ($C_{\varepsilon} \approx \mathrm{constant}$) and non-equilibrium ($C_{\varepsilon} \sim \mathrm{f}(Re_M)/Re_{\lambda}$) regular grid-generated decaying turbulence is presented. 

We have shown that the turbulent production and transport are large-scale effects which are negligible at length-scales smaller than $\ell/3$ even though our Reynolds numbers are moderate ($\ell$ is an integral length-scale taken, here, to be the longitudinal integral length-scale $L_{11}^{(1)}$). 
Hence, production and transport do not influence the maximum energy transfer to smaller-scales.

The maximum energy transfer rate $\Pi |_{\mathrm{max}}$ scales as $(\overline{u^{2}})^{3/2}/L_{11}^{(1)}$ both in the turbulence decay region which we term non-equilibrium region and in the further downstream turbulence decay region which we term equilibrium region.
The non-equilibrium region takes its name from the fact that $\varepsilon$ does not scale as $(\overline{u^{2}})^{3/2}/L_{11}^{(1)}$ in that region, thus indicating a severe scaling imbalance. 
In what we term the equilibrium region, $\Pi |_{\mathrm{max}}$ and $\varepsilon$ scale in
the same way. 
The imbalance between $\Pi |_{\mathrm{max}}$ and $\varepsilon$ in the non-equilibrium region drives the small-scale advection which is non-negligible and increases in proportion to the maximum energy transfer as the turbulence decays. 
Further downstream where the turbulence decay enters its equilibrium region, the small-scale advection remains about constant in proportion to the maximum energy transfer, presumably until the dissipation loses its high Reynolds number scaling $(\overline{u^{2}})^{3/2}/L_{11}^{(1)}$ because the local Reynolds number has decayed too much. 
However, we were not able to access such a very far downstream region in our experiments.

Finally, it should be stressed that the best defined power-law energy spectra with exponents closest to $-5/3$ in the grid-generated decaying turbulence have been recorded in the non-equilibrium region \cite[]{VV2012} where, irrespective of the fact that $-5/3$ is the Kolmogorov exponent, the lack of balance even in scaling terms between interscale transfer and dissipation indicates a clear non-Richardson-Kolmogorov cascade (see \citealp{MV2010}
and \citealp{VV2011}). 
It is important to know that non-equilibrium cascades such as the ones in the lee of various
grid-generated turbulent flows can follow well-defined scaling laws such as the one for dissipation studied in detail in \textbf{I} and the one for interscale energy transfer established here.\\

We are grateful to Prof. Arne Johansson (KTH) for the discussion concerning the experimental apparatus. 
P.C.V would like to thank Anthony R. Oxlade for the help in the digital imaging system setup and Ian Pardew and Roland Hutchins (aero workshop) for the manufacture of the apparatus. 
P.C.V. acknowledges the financial support from Funda\c{c}\~{a}o para a Ci\^{e}ncia e a Tecnologia (grant SFRH/BD/61223/2009, cofinanced by POPH/FSE).

\appendix

\section{Estimates of  $\Pi^*_U$ and $\mathcal{D}^*_{X,\nu}$} \label{sec:neglect}

\begin{figure}
\centering
\begin{lpic}{Figures/Thesis_NegligibleTerms(100mm)}
   \lbl[W]{-4,80,90;$-\Pi_{U}/\varepsilon$, $\mathcal{D}_{X,\nu}/\varepsilon$ ($\%$)}
   \lbl[W]{90,2;$r$ (mm)}
\end{lpic}
\caption{Negligible terms in \eqref{eq:KHM} averaged over spherical shells and normalised by the dissipation. (\protect\raisebox{-0.5ex}{\SmallCircle} $\!|\!\!$ \protect\raisebox{-0.5ex}{\SmallSquare})  $-\Pi_{U}/\varepsilon$ ($\%$) at $X_1=1250$mm and $X_1=2150$mm, respectively and (\protect\raisebox{-0.5ex}{\FilledSmallCircle} $\!|\!\!$ \protect\raisebox{-0.5ex}{\FilledSmallSquare}) $\mathcal{D}_{X,\nu}/\varepsilon$ ($\%$) at $X_1=1250$mm and $X_1=2150$mm, respectively. $X_1=1250$mm and $X_1=2150$mm correspond to $X_1/x_{\mathrm{peak}} = 1.5$ and $X_1/x_{\mathrm{peak}} = 2.6$, respectively.
Data are acquired in the lee of RG115.}
\label{fig:NegligibleTerms}
\end{figure}

The energy transfer due to mean velocity gradients, $\Pi^*_{U}$, and the transport via viscous diffusion, $\mathcal{D}^*_{X,\nu}$, are shown to be negligible compared to the other terms in \eqref{eq:KHM}. 
These terms are computed from the acquired data as described in \S \ref{sec:KHMcomputed}.
As shown in figure \ref{fig:NegligibleTerms} the term $\Pi_{U}^*$ averaged over spherical shells represents less than $0.4\%$ of the dissipation at $X_1/x_{\mathrm{peak}} = 1.5$ and further downstream, $X_1/x_{\mathrm{peak}} = 2.6$, it decreases to less than $0.05\%$.
The transport via viscous diffusion averaged over spherical shells, $\mathcal{D}_{X,\nu}$ is also negligible and represents less than $0.1\%$ for both downstream locations, $X_1/x_{\mathrm{peak}} = 1.5$ and  $X_1/x_{\mathrm{peak}} = 2.6$ (see figure \ref{fig:NegligibleTerms}).

\section{A kinematic upper bound for the
scale-by-scale viscous diffusion}\label{sec:appendix}

Under very plausible assumptions on the functional form of
\[S_{2}({\bf X}, {\bf r}) = \frac{1}{4\pi}\int d\Omega\,\overline{\delta q^{2}}({\bf X}, {\bf r})\] (the second-order structure function averaged over all directions ${\bf r}/r$ where $\Omega$ is
the solid angle)
we show that the spherical
averaged viscous diffusion term appearing in \eqref{eq:KHM} has an
upper bound of the form
\begin{equation}
\mathcal{D}_{\nu}(\mathbf{X},r)\equiv
\frac{\nu}{2}\frac{\partial^2\,S_2(\mathbf{X},r)}{\partial r_k^2}<
\frac{4\nu}{r^2} S_2(\mathbf{X},r)< \frac{4\nu}{r^2}
4K^{*}(\mathbf{X},r) \hspace{6mm} \forall r
\label{eq:appendix}
\end{equation}
where $4K^{*}(\mathbf{X},r)$ is the sum of twice the turbulent kinetic
energy at the two locations, $\mathbf{X} -\mathbf{r}/2$ and
$\mathbf{X} +\mathbf{r}/2$.

This inequality is useful in determining upper ranges of $r$ where
$\mathcal{D}_{\nu}(\mathbf{X},r)$ is negligible by comparison to some
other term in \eqref{eq:KHM}. Considering, for example, the
dissipation term $4\varepsilon^{*}(\mathbf{X},r)$ in \eqref{eq:KHM}, this
inequality can be used to show that if $\frac{4\nu}{r^2} 4K^{*}(\mathbf{X},r)\ll 4\varepsilon^{*}(\mathbf{X},r)$ then $\mathcal{D}_{\nu}(\mathbf{X},r) \ll 4\varepsilon^{*}(\mathbf{X},r)$. 
In other words, the spherical averaged viscous term $\mathcal{D}_{\nu}$
is neglible compared to $4\varepsilon^{*}$ in the upper range of scales
$r \gg \lambda^{*}$ where $\lambda^{*} \equiv \sqrt{4\nu
K^{*}/\varepsilon^{*}}$. 
Clearly $\lambda^*$ is close to a fraction of the Taylor microscale $\lambda$, in fact close to 
$\sqrt{2/5} \lambda$, where turbulent kinetic energy gradients and
turbulent dissipation gradients are small and this is indeed the case in the flow regions where the results reported in \S \ref{sec:5} are observed. 
One of these results is that $\mathcal{D}_{\nu}$ is small compared to $4\varepsilon^{*}$ for $r> \lambda$ (see figures 10b and 11b), a result which can therefore be considered to be a simple kinematic consequence of the inequality established in this Appendix.
This conclusion and \eqref{eq:appendix} in general are generalisations in physical space of similar results previously obtained in Fourier space for homogeneous turbulence by a very
different method \cite[]{Laizet2013}.

We now proceed by proving the inequalities \eqref{eq:appendix}.

We start by writing the Laplacian of $S_2(\mathbf{X},r)$ in spherical
coordinates,
\[\frac{\partial^2 \,S_2(\mathbf{X},r)}{\partial r_k^2}=r^{-2}
\frac{\partial}{\partial r}\left( r^2 \frac{\partial
S_2(\mathbf{X},r)}{\partial r} \right)= \frac{\partial^{2}
S_2(\mathbf{X},r)}{\partial r^{2}} + \frac{2}{r}\frac{\partial
S_2(\mathbf{X},r)}{\partial r}\]
and noting that
\[\lim_{r \rightarrow 0}\frac{\partial^2 \,S_2(\mathbf{X},r)}{\partial
r_k^2} = 3\,\lim_{r \rightarrow 0}
\frac{\partial^2\,S_2(\mathbf{X},r)}{\partial r^2} =
\varepsilon(\mathbf{X})/\nu .\]
Taylor expanding about $r=0$ implies that $S_2(\mathbf{X},r) = 3\,\varepsilon(\mathbf{X})/(2\,\nu)\, r^2$ for small enough {values} of $r$. 
At large enough values of $r$, $S_2(\mathbf{X},r)\approx
4K^{*}(\mathbf{X},r)$; more accurately, $\lim_{r \rightarrow \infty} S_2 (\mathbf{X},r)= 4K^{*}(\mathbf{X},r)$

We assume (i) that $S_2$ is a monotonically increasing function in
$0\leq r < \infty$ with continuous first- and second-order derivatives
with respect to $r$; (ii) that it has only one inflection point at $r
= r_{I}$, i.e. $\frac{\partial^2 S_2(\mathbf{X},r)}{\partial r^2} (r)=0$
only at $r=r_I$; and (iii) that $\frac{\partial
S_2(\mathbf{X},r)}{\partial r} (r)$ is concave in the range $0\le r
\le r_{I}$. The monotonicity assumption directly implies that $S_{2} <
4K^{*}$ which deals with the second inequality in \eqref{eq:appendix}.

The existence of an inflection point is consistent with the commonly
observed functional form of $S_2(r)$ where $S_2(\mathbf{X},r) \sim
r^2$ for small r followed by a smooth transition to a power-law of the
type $S_2(\mathbf{X},r)\sim r^n$ with $n<1$ ($n=2/3$ for Kolmogorov's
inertial range). The inflection point resides at a value of $r$
between these two power laws. The absence of another inflection point
agrees with the monotonically increasing passage from the $n<1$ power
law to a constant (independent of $r$).

The assumption that $\frac{\partial S_2(\mathbf{X},r)}{\partial r}
(r)$ is concave in the range $0\le r \le r_{I}$ and then monotonically
decreasing at $r>r_I$ is enough to establish that
\begin{equation}
S_2(\mathbf{X},r)=\int_{0}^{r} \frac{\partial
S_2(\mathbf{X},\zeta)}{\partial \zeta}\,d\zeta \,\,>\,\, \frac{r}{2}
\frac{\partial S_2(\mathbf{X},r)}{\partial r}\hspace{8mm} \forall r.
\label{A2}
\end{equation}
This can be seen as a geometrical inequality relating the area of the
triangle of base $r$ and height $\partial S_2(\mathbf{X},r)/\partial r$ with
the area underneath $\partial S_2(\mathbf{X},\zeta)/\partial \zeta$ for
$0\leq\zeta\leq r$.
By a similar geometric reasoning for $r\le r_I$,
\begin{equation}
\frac{\partial S_2(\mathbf{X},r)}{\partial r}=\int_{0}^{r} \frac{\partial^2
S_2(\mathbf{X},\zeta)}{\partial \zeta^2}\,d\zeta \,\,>\,\, \frac{r}{2}
\frac{\partial^2 S_2(\mathbf{X},r)}{\partial r^2}\hspace{8mm} \forall r\leq
r_I
\label{A3}
\end{equation}
which combined with  \eqref{A2} leads to
\begin{equation}
r^{-2} \frac{\partial}{\partial r}\left( r^2 \frac{\partial
S_2(\mathbf{X},r)}{\partial r}\right) =  \frac{\partial^2
S_2(\mathbf{X},r)}{\partial r^2} + \frac{2}{r}\frac{\partial
S_2(\mathbf{X},r)}{\partial r} < \frac{8 S_2(\mathbf{X},r)}{r^2}
\hspace{4mm} \forall r\leq r_I.
\label{A4}
\end{equation}

For $r>r_{I}$, $\partial^2 S_2(\mathbf{X},r)/\partial r^2<0$ and
$\partial S_2(\mathbf{X},r)/\partial r>0$ by assumption, thus
\begin{equation}
\frac{\alpha}{r} \frac{\partial S_2(\mathbf{X},r)}{\partial
r}>\frac{\partial^2 S_2(\mathbf{X},r)}{\partial r^2}
\hspace{12mm} \forall \alpha\geq 0,
\label{A5}
\end{equation}
which  together with  \eqref{A2} leads to
\begin{equation}
r^{-2} \frac{\partial}{\partial r}\left( r^2 \frac{\partial
S_2(\mathbf{X},r)}{\partial r}\right)  < \frac{2(2+\alpha)
S_2(\mathbf{X},r)}{r^2} \hspace{4mm} \forall \alpha\geq0,\, \forall r\geq
r_I.
\label{A6}
\end{equation}
Inequalities \eqref{A4} and \eqref{A6} (with $\alpha=2$ for
convenience) can now be combined to yield the first inequality in
\eqref{eq:appendix}.

\section{Equality of $\Pi |_{\mathrm{max}}$ in physical and wavenumber space}\label{sec:appendixB}

We note that the maximum nonlinear energy transfer in physical space,  $\Pi |_{\mathrm{max}}$, is equal to its wavenumber space counterpart, $\Pi_K |_{\mathrm{max}}=\Pi |_{\mathrm{max}}$ ($\Pi_K \equiv \int_0^K\! T(k)\,dk$, where $T(k)$ is the spherical averaged non-linear spectral transfer term, see e.g. \citealt{Frisch:book}). 
This can be seen from (6.17) of \cite{Frisch:book}, noting that $\boldsymbol{\nabla_{\ell}}\cdot(\boldsymbol{\ell}/\ell^2 \,\,\Pi(\ell)\,)|_{\Pi(\ell) = \Pi |_{\mathrm{max}}} =  \Pi |_{\mathrm{max}} /\ell^2$ and that $\int_{\mathbb{R}^3} d^3\ell\, \mathrm{sin}(K\ell)/\ell^3 = 2\pi^2$ (using the book's notation and defining $\Pi(\ell) \equiv \boldsymbol{\nabla_\ell} \cdot \left<|\mathbf{\delta u}(\boldsymbol{\ell})|^2\mathbf{\delta u}(\boldsymbol{\ell})\right>/4$ and $\Pi |_{\mathrm{max}} \equiv \mathrm{max}\,(|\Pi(\ell)|)$). For a inhomogeneous turbulent flow, care must be taken in guaranteeing existence of  the Fourier transform \cite[]{Deissler61,Deissler81}.

Contrastingly, it is not straightforward to establish a simple relationship between $\overline{\delta u_{\parallel}\delta q^2}/r$ or $\overline{(\delta u_{\parallel})^3}/r$ and a wavenumber space counterpart,  see discussion in \S IV of \cite{Cambon2012}.

\bibliographystyle{plainnat}
\bibliography{mybib}

\begin{thebibliography}{44}
\providecommand{\natexlab}[1]{#1}
\providecommand{\url}[1]{\texttt{#1}}
\expandafter\ifx\csname urlstyle\endcsname\relax
  \providecommand{\doi}[1]{doi: #1}\else
  \providecommand{\doi}{doi: \begingroup \urlstyle{rm}\Url}\fi

\bibitem[Antonia and Burattini(2006)]{Antonia2006}
R.~A. Antonia and P.~Burattini.
\newblock Approach to the 4/5 law in homogeneous isotropic turbulence.
\newblock \emph{J. Fluid Mech.}, 550:\penalty0 175--184, 2006.

\bibitem[Benedict and Gould(1996)]{BG96}
L.~H. Benedict and R.~D. Gould.
\newblock Towards better uncertainty estimates for turbulent statistics.
\newblock \emph{Exp. Fluids}, 22:\penalty0 129--136, 1996.

\bibitem[Biskamp(2003)]{Biskamp:book}
D.~Biskamp.
\newblock \emph{Magnetohydrodynamic Turbulence}.
\newblock Cambridge University Press, 2003.

\bibitem[Borue and Orszag(1998)]{BO98}
V~Borue and S.~A. Orszag.
\newblock Local energy flux and subgrid-scale statistics in three-dimensional
  turbulence.
\newblock \emph{J. Fluid Mech.}, 366\penalty0 (1 -- 31), 1998.

\bibitem[Bos et~al.(2007)Bos, Shao, and Bertoglio]{Bos2007}
W.~J.~T. Bos, L.~Shao, and J.-P. Bertoglio.
\newblock Spectral imbalance and the normalized dissipation rate of turbulence.
\newblock \emph{Phys. Fluids}, 19\penalty0 (045101), 2007.

\bibitem[Danaila et~al.(2012)Danaila, Krawczynski, Thiesset, and
  Renou]{Danaila2012}
L.~Danaila, J.~F. Krawczynski, F~Thiesset, and B.~Renou.
\newblock Yaglom-like equation in axisymmetric anisotropic turbulence.
\newblock \emph{Physica D}, 241\penalty0 (3):\penalty0 216--223, 2012.

\bibitem[{de Gennes}(1990)]{deGennes:book}
P.~G. {de Gennes}.
\newblock \emph{Introduction to polymer dynamics}.
\newblock Cambridge University Press, 1990.

\bibitem[Deissler(1961)]{Deissler61}
R.~G. Deissler.
\newblock Effects of inhomogeneity and of shear flow in weak turbulent fields.
\newblock \emph{Phys. Fluids}, 4\penalty0 (1187), 1961.

\bibitem[Deissler(1981)]{Deissler81}
R.~G. Deissler.
\newblock Spectral energy transfer for inhomogeneous turbulence.
\newblock \emph{Phys. Fluids}, 24\penalty0 (1911), 1981.

\bibitem[Discetti et~al.(2013)Discetti, Ziskin, Astarita, Adrian, and
  Prestridge]{discettietal11}
S.~Discetti, I.~B. Ziskin, T.~Astarita, R.~J. Adrian, and K.~P. Prestridge.
\newblock {PIV} measurements of anisotropy and inhomogeneity in decaying
  fractal generated turbulence.
\newblock \emph{Fluid Dyn. Res.}, {\rm Accepted for publication}, 2013.

\bibitem[Frisch(1995)]{Frisch:book}
U.~Frisch.
\newblock \emph{{Turbulence: The Legacy of AN Kolmogorov}}.
\newblock Cambridge University Press, Cambridge, 1995.

\bibitem[Gagne et~al.(2004)Gagne, Castaing, Baudet, and Malecot]{Gagne2004}
Y.~Gagne, B.~Castaing, C.~Baudet, and Y.~Malecot.
\newblock Reynolds dependence of third-order structure functions.
\newblock \emph{Phys. Fluids}, 16\penalty0 (2):\penalty0 482, 2004.

\bibitem[George(2013)]{George2013}
W.~K. George.
\newblock Reconsidering the {`Local Equilibrium'} hypothesis for small scale
  turbulence.
\newblock \emph{Proceedings Marseille (to appear)}, 2013.

\bibitem[Gomes-Fernandes et~al.(2012)Gomes-Fernandes, Ganapathisubramani, and
  Vassilicos]{gomesfernandesetal12}
R.~Gomes-Fernandes, B.~Ganapathisubramani, and J.~C. Vassilicos.
\newblock Particle image velocimetry study of fractal-generated turbulence.
\newblock \emph{J. Fluid Mech.}, 711:\penalty0 306--336, 2012.

\bibitem[K\'{a}rm\'{a}n and Howarth(1938)]{KH1938}
T.~K\'{a}rm\'{a}n and L.~Howarth.
\newblock {On the statistical theory of isotropic turbulence}.
\newblock \emph{Proc. Roy. Soc. A}, 164\penalty0 (917):\penalty0 192--215,
  1938.

\bibitem[Kendall and Stuart(1958)]{KS58}
M.~G. Kendall and A.~Stuart.
\newblock \emph{The advanced theory of statistics}, volume~1.
\newblock {Charles Griffin \& Co Limited, London}, 1958.

\bibitem[Kolmogorov(1941a)]{K41a}
A.~N. Kolmogorov.
\newblock The local structure of turbulence in incompressible viscous fluid for
  very large reynolds numbers.
\newblock \emph{Dokl. Akad. Nauk. SSSR}, 30\penalty0 (4), 1941a.
\newblock (Reprinted in 1991, Proc. R. Soc. Lond., Vol. 434, pp. 9-13).

\bibitem[Kolmogorov(1941c)]{K41c}
A.~N. Kolmogorov.
\newblock Dissipation of energy in the locally isotropic turbulence.
\newblock \emph{Dokl. Akad. Nauk. SSSR}, 32\penalty0 (1), 1941c.
\newblock (Reprinted in 1991, paper 47 in Selected Works of A.N. Kolmogorov,
  Vol. I: Mathematics and Mechanics).

\bibitem[Kraichnan(1974)]{Kraichnan74}
R.~Kraichnan.
\newblock On {Kolmogorov's} inertial-range theories.
\newblock \emph{J. Fluid Mech.}, 62(2):\penalty0 305 -- 330, 1974.

\bibitem[Laizet and Vassilicos(2011)]{Sylvain2011}
S.~Laizet and J.~C. Vassilicos.
\newblock {DNS} of fractal-generated turbulence.
\newblock \emph{Flow Turb. Combust.}, pages 87:673--705, 2011.

\bibitem[Laizet et~al.(2013)Laizet, Vassilicos, and Cambon]{Laizet2013}
S.~Laizet, J.~C. Vassilicos, and C.~Cambon.
\newblock Interscale energy transfer in decaying turbulence and
  vorticity-strain dynamics in grid-generated turbulence.
\newblock \emph{Fluid Dyn. Res. (to appear October 2013)}, 2013.

\bibitem[Lamriben et~al.(2011)Lamriben, Cortet, and Moisy]{Moisy2011}
C.~Lamriben, P.-P. Cortet, and F.~Moisy.
\newblock Direct measurements of anisotropic energy transfers in a rotating
  turbulence experiment.
\newblock \emph{Phys. Rev. Lett.}, 107\penalty0 (024503), 2011.

\bibitem[Lumley(1992)]{Lumley92}
J.~L. Lumley.
\newblock Some comments on turbulence.
\newblock \emph{Phys. Fluids A}, 4\penalty0 (2):\penalty0 203--211, 1992.

\bibitem[Lundgren(2002)]{Lundgren2002}
T.~S. Lundgren.
\newblock Kolmogorov two-thirds law by matched asymptotic expansion.
\newblock \emph{Phys. Fluids}, 14\penalty0 (2), 2002.

\bibitem[Lundgren(2003)]{Lundgren2003}
T.~S. Lundgren.
\newblock Kolmogorov turbulence by matched asymptotic expansion.
\newblock \emph{Phys. Fluids}, 15\penalty0 (4), 2003.

\bibitem[Marati et~al.(2004)Marati, Casciola, and Piva]{Casciola04}
N.~Marati, C.~M. Casciola, and R.~Piva.
\newblock Energy cascade and spatial fluxes in wall turbulence.
\newblock \emph{J. Fluid Mech.}, 521:\penalty0 191--215, 2004.

\bibitem[Mazellier and Vassilicos(2010)]{MV2010}
N.~Mazellier and J.~C. Vassilicos.
\newblock Turbulence without {Richardson-Kolmogorov} cascade.
\newblock \emph{Phys. Fluids}, 22:\penalty0 075101, 2010.

\bibitem[McComb et~al.(2010)McComb, Berera, Salewski, and Yoffe]{McComb2010}
W.~D. McComb, A.~Berera, M.~Salewski, and S.~Yoffe.
\newblock Taylor's (1935) dissipation surrogate reinterpreted.
\newblock \emph{Phys. Fluids}, 22\penalty0 (061704), 2010.

\bibitem[Moisy et~al.(1999)Moisy, Tabeling, and Wilaime]{Moisy1999}
F.~Moisy, P.~Tabeling, and H.~Wilaime.
\newblock Kolmogorov equation in a fully developed turbulence experiment.
\newblock \emph{Phys. Rev. E}, 82\penalty0 (20):\penalty0 3994, 1999.

\bibitem[Monin and Yaglom(1975)]{MY75}
A.S. Monin and A.M. Yaglom.
\newblock \emph{Statistical Fluid Mechanics}, volume Vol. 2.
\newblock MIT Press, 1975.

\bibitem[Nagata et~al.(2013)Nagata, Sakai, Inaba, Suzuki, Terashima, and
  Suzuki]{Nagata2012}
K.~Nagata, Y.~Sakai, T.~Inaba, H.~Suzuki, O.~Terashima, and H.~Suzuki.
\newblock Turbulence structure and turbulence kinetic energy transport in
  multiscale/fractal-generated turbulence.
\newblock \emph{Phys. Fluids}, 25\penalty0 (065102), 2013.

\bibitem[Nie and Tanveer(1999)]{NT99}
Q.~Nie and S.~Tanveer.
\newblock A note on third-order structure functions in turbulence.
\newblock \emph{Proc. R. Soc. Lond. A}, 455:\penalty0 1615--1635, 1999.

\bibitem[Qian(1999)]{Qian99}
J.~Qian.
\newblock Slow decay of the finite reynolds number effect of turbulence.
\newblock \emph{Phys. Rev. E}, 60\penalty0 (3), 1999.

\bibitem[Rubinstein and Bos(2009)]{RB2009}
R.~Rubinstein and W.~J.~T. Bos.
\newblock On the unsteady behavior of turbulence models.
\newblock \emph{Phys. Fluids}, 21\penalty0 (041701), 2009.

\bibitem[Schiestel(1987)]{Schiestel87}
R.~Schiestel.
\newblock Multiple-time-scale modeling of turbulent flows in one point
  closures.
\newblock \emph{Phys. Fluids}, 30\penalty0 (722), 1987.

\bibitem[Singh and Bhadauria(2009)]{SB09}
A.~K. Singh and B.~S. Bhadauria.
\newblock Finite difference formulae for unequal sub-intervals using
  {L}agrange's interpolation formula.
\newblock \emph{Int. J. Math. Anal.}, 3\penalty0 (17), 2009.

\bibitem[Sj\"{o}gren(1997)]{SThesis}
T.~Sj\"{o}gren.
\newblock \emph{Development and Validation of Turbulence Models Through
  Experiment and Computation}.
\newblock PhD thesis, Royal Institute of Technology (K.T.H.), 1997.

\bibitem[Sj\"{o}gren and Johansson(1998)]{SJ98}
T.~Sj\"{o}gren and A.~V. Johansson.
\newblock Measurement and modelling of homogeneous axisymmetric turbulence.
\newblock \emph{J. Fluid Mech.}, 374:\penalty0 59--90, 1998.

\bibitem[Tchoufag et~al.(2012)Tchoufag, Sagaut, and Cambon]{Cambon2012}
J.~Tchoufag, P.~Sagaut, and C.~Cambon.
\newblock Spectral approach to finite {R}eynolds number effects on
  {K}olmogorov's 4/5 law in isotropic turbulence.
\newblock \emph{Phys. Fluids}, 24\penalty0 (015107), 2012.

\bibitem[Valente and Vassilicos(2011)]{VV2011}
P.~C. Valente and J.~C. Vassilicos.
\newblock The decay of turbulence generated by a class of multi-scale grids.
\newblock \emph{J. Fluid Mech.}, 687:\penalty0 300--340, 2011.

\bibitem[Valente and Vassilicos(2012)]{VV2012}
P.~C. Valente and J.~C. Vassilicos.
\newblock Universal dissipation scaling for nonequilibrium turbulence.
\newblock \emph{Phys. Rev. Lett.}, 108\penalty0 (214503), 2012.

\bibitem[Valente and Vassilicos(2014)]{VV2013}
P.~C. Valente and J.~C. Vassilicos.
\newblock {The nonequilibrium region of grid-generated turbulence}.
\newblock \emph{J. Fluid Mech. (in press)}, 2014.

\bibitem[Wan et~al.(2010)Wan, Xiao, Meneveau, Eyink, and Chen]{MY2010}
M.~Wan, Z.~Xiao, C.~Meneveau, G.~L. Eyink, and S.~Chen.
\newblock Dissipation-energy flux correlations as evidence for the lagrangian
  energy cascade in turbulence.
\newblock \emph{Phys. Fluids}, 22\penalty0 (061702), 2010.

\bibitem[Yoshizawa(1994)]{Yoshizawa1994}
A.~Yoshizawa.
\newblock Nonequilibrium effect of the turbulent-energy-production process on
  the inertial-range spectrum.
\newblock \emph{Phys. Rev. E}, 49\penalty0 (5), 1994.

\end{thebibliography}

\end{document}